\documentclass[draft]{agujournal2019}
\usepackage{url} 
\usepackage{soul}

\usepackage{natbib}
\usepackage{tabularx}
\usepackage{array}
\usepackage{hhline}
\usepackage{booktabs}
\usepackage{hyperref}

\usepackage{float} 
\usepackage{amsmath}
\usepackage{rotating}

\usepackage{ragged2e}
\justifying

\usepackage{booktabs}
\usepackage{multirow}
\usepackage{makecell}
\usepackage{subcaption}
\draftfalse

\journalname{JGR: Planets}

\begin{document}

%
%


\title{Thermal Segregation and Reddening in Europa's Double Ridges}

%
%




\authors{Kya C. Sorli\affil{1,2}\thanks{https://orcid.org/0009-0009-0386-6241}, Paul O. Hayne\affil{1}\thanks{https://orcid.org/0000-0003-4399-0449}, Lucas Lange\affil{3}\thanks{https://orcid.org/0000-0001-6433-1050}, Sylvain Piqueux\affil{3}\thanks{https://orcid.org/0000-0003-0485-2908}}


\affiliation{1}{Department of Astrophysical \& Planetary Sciences, Laboratory for Atmospheric \& Space Physics, University of Colorado Boulder, Boulder, CO, USA} 
\affiliation{2} {Department of Physics, University of Oxford, Oxford, UK }
\affiliation{3}{Jet Propulsion Laboratory, California Institute of Technology, Pasadena, CA, USA}





\correspondingauthor{Kya C. Sorli}{kya.sorli@colorado.edu}



\begin{keypoints}
\item We use a 3D thermophysical model to investigate whether thermal segregation can produce observable reddening in Europa's double ridges
\item Self-heating in ridge troughs leads to higher maximum temperatures and sublimation rates compared to surrounding terrain
\item Thermal segregation can produce dark lag layers from the equator to roughly 60 degrees latitude, depending on exospheric density 
\end{keypoints}

%
%

%
%


\begin{abstract}
   
     Europa's double ridges often display lower albedo and redder color than their surroundings. Their unique topography may cause sublimation-driven darkening due to illumination and self-heating -- the process of thermal segregation. We apply an advanced 3D thermophysical model, including shadowing and self-heating through mutual exchange of radiation, to digital elevation models of double ridges at a range of latitudes and orientations. Results show that self-heating in ridge troughs can markedly increase temperatures and sublimation rates, with a difference in maximum trough temperatures of up to 20 K, which may have implications for detection of endogenic heat. Incorporating a simple exosphere model and assuming an initial 10\% concentration of 1 $\mu$m non-ice particles, we find thermal segregation can produce reddening in the form of dark lag layers from the equator to the middle latitudes, but is generally negligible at 60 degrees or higher. Lag formation timescales in ridge troughs are 10 - 100 yr to produce an optically thick layer. Modeling suggests that low-albedo lag layer formation provides positive feedback, further increasing surface heating. These effects may also darken Europa's surface in areas surrounding the ridges. However, the net mass balance controlling sublimation and lag formation is highly sensitive to the global water exosphere density: values $\sim 10^{16}$ molec/m$^{2}$ produce reddening in the trough and ablation of $\sim1~\mu\mathrm{m~yr^{-1}}$ of material, while values $\sim10^{18}$ molec/m$^{2}$ result in net deposition of $\sim 10~\mu\mathrm{m~yr^{-1}}$. Model predictions of resulting low albedo material in double ridge troughs are provided, which can be tested with eventual data from Europa Clipper.
    
\end{abstract}

\section*{Plain Language Summary}
Jupiter's moon Europa is covered with unique landscapes. These include distinctive double ridges, where two raised ridges surround a large central trough. Many double ridges are covered with dark or reddened material, which may be from the ice shell or the subsurface ocean. We investigate whether this reddening may come from another source. Increased temperatures from topography can cause accelerated ice loss and concentration of dark material in a process known as thermal segregation. We use a 3D thermophysical model that includes heating from topography to estimate temperatures in double ridges at different latitudes across Europa, and calculate the associated ice loss. Model results suggest that troughs can reach higher temperatures than their surroundings, which may have implications for finding evidence of hot spots or plume activity. Incorporating a simple exosphere model, we predict that observable lag layers may form as quickly as 10 - 100 years but that the polar regions should remain cold. Once dark areas form, they can speed the reddening of surrounding terrain. This mechanism may be responsible for some of the darkening on Europa's surface, and may interact with other processes. These results have implications for the data that will eventually be returned by Europa Clipper.

%
%

%


%
%
%
%

\section{Introduction and Background}

Following its discovery in 1610, observations as early as the 1950s suggested that Europa was a bright, potentially icy world \citep{Kuiper1957InfraredSatellites}. Later visits by spacecraft, including Pioneer, Voyager 1 and 2, and Galileo, revealed a nuanced surface. Now widely accepted as an ocean world \citep{Kivelson2000GalileoEuropa}, Europa's surface is comparatively young and sparsely cratered but hosts unique features such as lineae, chaos terrain, and ridged plains. The latter of these are marked by tectonic features, including the ubiquitous ``double ridges" (e.g. \citet{Kattenhorn2009TectonicsEuropa, Prockter2009MorphologyBands}). Found globally across Europa, these double ridges are common and often crosscut each other or other terrain features. Composed of two ridges, or raised ice, with a central trough between, they are usually of order 100 meters high, about five kilometers across, and can extend hundreds of kilometers in length \citep{Collins2009ChaoticEuropa}. Due to uncertainties regarding Europa's ice shell and interior \citep{Roberts2023ExploringClipper}, the formation mechanism for double ridges is unresolved \citep{Daubar2024PlannedMission}. Theories include squeezing from tidal stresses \citep{Greenberg1998TectonicFeatures}, shear heating \citep{Han2008ImplicationsEuropa}, crystallization of a liquid water intrusion \citep{Johnston2014FormationShell}, cryovolcanism \citep{Kadel1998Trough-BoundingFormation}, or formation via incremental ice wedging \citep{Cashion2024EuropasWedging}. 

\begin{figure}
    \centering
    \includegraphics[width=0.99\linewidth]{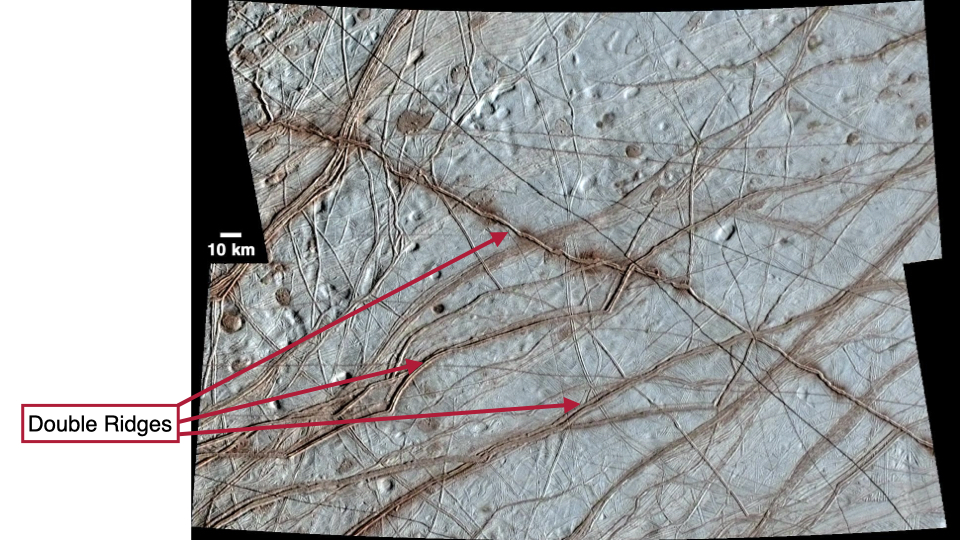}
    \caption{An image taken by the Solid State Imager onboard the Galileo spacecraft showing the Rhadamanthys region, including its lineae and varied topography. The Rhadamanthys Lineae offer several examples of prominent double ridges, a few examples of which are labeled. Low albedo material is closely correlated with the topography for many ridges. For others, including the ridge cutting from northwest to southeast, reddening is more diffuse and can be found on the ridge margins. Note the pits in the top left of the image, which can also display varied amounts of reddening. Image from NASA/JPL-Caltech/University of Arizona.}
    \label{fig:doubleridges}
\end{figure}

Images from the NASA Galileo mission introduced new questions regarding the double ridges, as many exhibit distinctly lower albedo than their surroundings. Dark, reddish material is often visible and concentrated along the central troughs and on the tops of ridges \citep{McCord2010HydratedInvestigation}. For some ridges, the reddened material is tightly aligned with the ridge itself, while for others material is distributed in diffuse patterns spread around the ridge margins. This diffuse material can extend past the margins of ridges, or be found in circular deposits around the lineae. In addition, some troughs display a significantly lower albedo than others. Though this is likely the product of multiple competing factors, it is impacted by the thickness of the dark layer as well as well as differing composition, which has a strong effect on visual albedo on Europa (e.g. \cite{Carlson2005DistributionHydrate}, \cite{Ligier2016VLT/SINFONICOMPOSITION}). Using Galileo NIMS data, \cite{McCord1999HydratedInvestigation} found that this dark material is non-ice, with spectral signatures matching hydrated Na- and Mg- salts or sulfuric acid hydrates \citep{Carlson1999SulfuricCycle}. Examples of prominent crisscrossing double ridges can be found in the Rhadamanthys lineae (Fig. \ref{fig:doubleridges}). Reddening can also be observed to varying degrees on other geological features in Fig. \ref{fig:doubleridges}, including ridged dilation bands and the pitted lenticulae.

Determining the cause of reddening in Europa's ridged plains is crucial to understanding the processes reshaping Europa's surface and may help to gain insight into the moon's interior. Several theories have been proposed. \citet{Nimmo2002Strike-slipEuropa} suggested frictional heating and advection along fault lines may concentrate non-ice material via ice sublimation at Europa's surface. In contrast, works such as \cite{Pappalardo1998GeologicalShell} and \citet{McCord2010HydratedInvestigation} proposed that thermal or compositional buoyancy in the ice shell may induce upwelling of non-ice material. Deposition of dark material arising from a subsurface ocean may also result from plume activity \citep{Quick2020CharacterizingEuropa}. However, another process known as thermal segregation has also been suggested \citep{Spencer1987ThermalSatellites}. 

\begin{figure}
    \centering
    \includegraphics[width=0.9\linewidth]{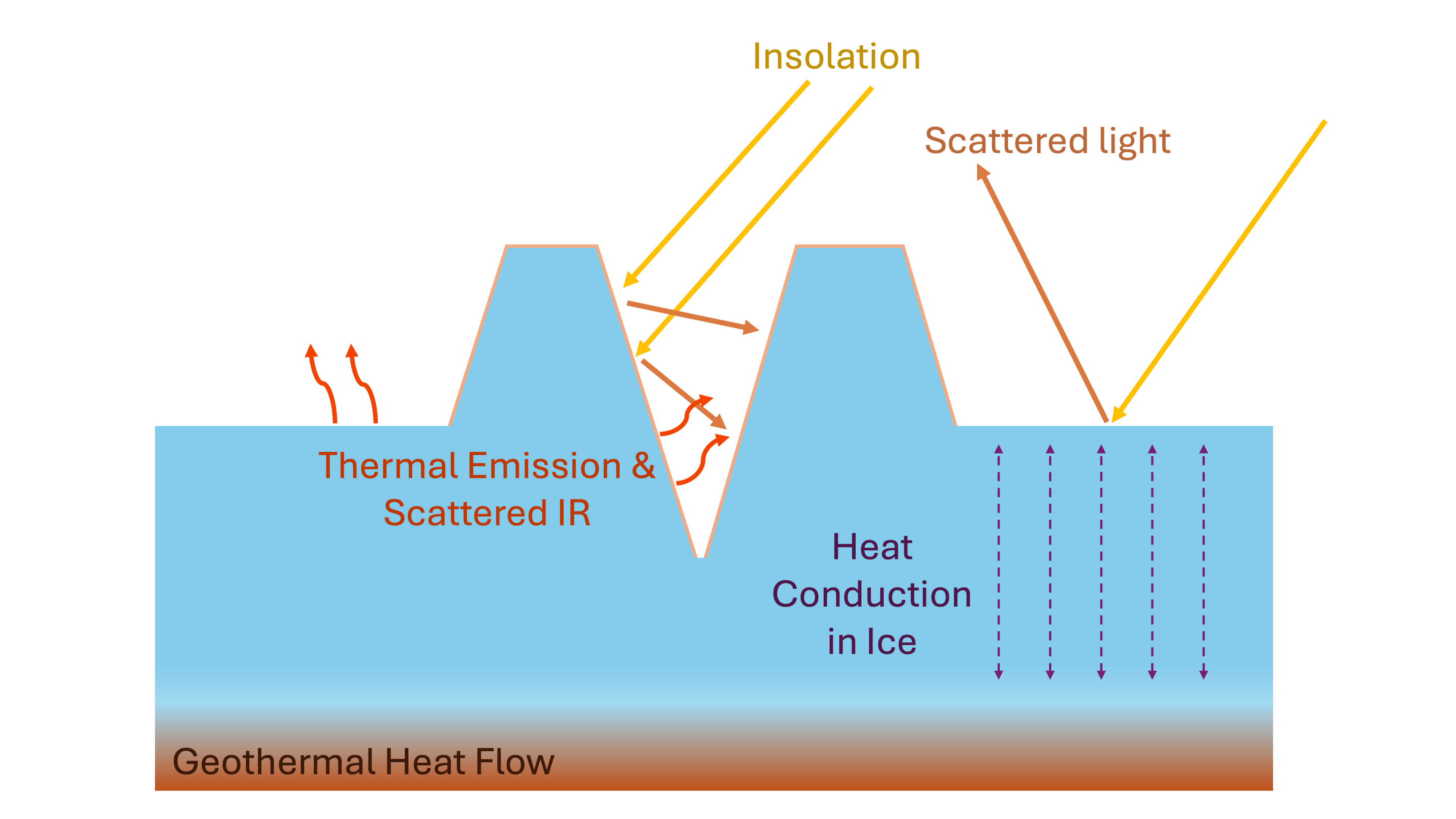}
    \caption{Diagram illustrating how topography can lead to temperature contrasts. Concave features can result in significant self-heating due to reflected insolation or absorption of thermal re-emission. This results in higher temperatures and higher sublimation rates, leading to thermal segregation of materials.}
    \label{fig:tsdiagram}
\end{figure}

On icy worlds, the composition of the ice itself can significantly alter surface temperatures. Thermal segregation occurs when dark material or topography causes locally hotter temperatures that produce positive temperature feedback. For low albedo regions, darker non-ice material will be warmer than higher albedo material around it, resulting in increased ice loss from the area or surroundings. Topography can also lead to locally hotter areas, such as on sun-facing slopes or in concave formations that receive secondary heating. This is illustrated in Figure \ref{fig:tsdiagram}. In these concavities, topography can enhance temperature contrasts, leading to more rapid sublimation of material and concentration of non-ice material, which in turn leads to warmer temperatures due to lower albedos. Higher temperatures also raise ice migration rates and sputtering yields \citep{Brown1982FastImplications}, thus having implications for the density and composition of the water exosphere \citep{Kumar1982TheSatellites}). Thermal segregation is thought to play a role in reshaping the surfaces of many outer solar system icy worlds, and has been suggested for Galilean satellites previously. \citet{Spencer1987ThermalSatellites} modeled thermal segregation on the Galilean satellites and found sublimation could be a powerful mechanism for ice redistribution on sub-km scales. \citet{Prockter1998DarkResolution} found that thermal segregation may provide an explanation for low albedo material on ridge slopes on Ganymede. Extreme cases, such as the albedo dichotomy on Iapetus, may be due to dark material causing runaway global thermal water ice migration \citep{Spencer2010FormationMigration}.

Though thermal segregation could be an active process on icy worlds, its efficacy at darkening Europa's surface depends on several competing processes, including sputtering and the density of the water exosphere. At high temperatures, sublimation serves as a principal loss mechanism for icy surfaces. At lower temperatures, sputtering by ions and electrons dominate. In turn, the exosphere is maintained and dependent on these same processes. The Jovian magnetosphere produces strong space weathering on Europa's surface, leading to distinctive large scale alteration of the surface and hemispherical differences \citep{Cassidy2013MagnetosphericEuropa,Pospieszalska1989MagnetosphericDione}. Due to bombardment of magnetospheric flow of ions from the Io-Europa plasma torus, Europa's trailing hemisphere is markedly darker and redder than its leading hemisphere \citep{Hendrix2011EuropasTerrains}. This bombardment can also result in net ablation of material, through a process of ion erosion or sputtering. 

Understanding the effect of thermal processes on Europa's surface is especially timely due to anticipated data from NASA's Europa Clipper mission \citep{Pappalardo2024ScienceMission}.  Launched in October of 2024, Europa Clipper is carrying the Europa Thermal Imaging System, (E-THEMIS), which will map 80\% of Europa with thermal infrared images at a resolution of $\le$ 8 km per pixel \citep{Christensen2024TheMission}. E-THEMIS will investigate areas of geologic activity and the formation of surface features, as well as search for future landing sites. Thermophysical models can be leveraged to make predictions about evolutionary and formation processes active on Europa's surface that will be directly testable upon Clipper's arrival at Europa in 2030. 

In addition, thermophysical models are necessary to interpret future E-THEMIS data, and notably for the identification of surface thermal anomalies that may originate from endogenic-heat. This is a key objective of the Europa Clipper mission \citep{Pappalardo2024ScienceMission}. Experience with Enceladus \citep{Spencer2006CassiniSpot} and signs of recent surface activity on Europa \citep{Schmidt2011ActiveEuropa, Roth2014TransientPole} suggest that complex topographic constructs, like ridges and chaos on Europa, may represent some of the most promising terrains where putative hot spots could preferentially be located. Lessons learned from past missions throughout the solar system indicate that seemingly anomalous warm surfaces are often associated with complex topography, yielding complex illumination patterns and self heating that could mistakenly be interpreted as low intensity hot spots. For instance, collapsed lava tubes and steep canyons on Mars appear 10-20 K warmer than their surroundings. This is partially due to self-heating between the canyon’s walls and reduced sky- view, limiting nighttime cooling, and possibly because of coarser scarp material with a high thermal inertia which retain daytime heating during night \citep{Christensen2003MorphologyResults}. Similarly, lunar pits can also appear 10's of K warmer than their surroundings, again due to self-heating \citep{Horvath2022ThermalExperiment}.

Thermophysical modeling can be applied to test where dark lag material may form, and whether it can explain observed patterns on the moon's surface. We begin by estimating the magnitude of the self heating effect using analytical approximation in 2D. However, thermal segregation and its contributions to the darkening/reddening of double ridges is intrinsically a 3D problem due to the effects of topography on insolation, shadowing, and self-heating in topographic depressions. Thus, we develop a 3D ray tracing thermophysical model capable of modeling the thermal segregation process, which includes surface heating due to secondary radiation. The model incorporates digital elevation models (DEMs) derived from Galileo image data to investigate whether thermal segregation can produce observable reddening in double ridge formations. By calculating sublimation rates at different locations using the temperature calculations, we also calculate timescales for formation of optically thick non-ice lag layers due to ablation of icy material containing impurities. Finally, we estimate how the formation rates and surface mass balance of these low albedo lag layers are affected by global processes such as temperature-dependent sputtering and deposition from the exosphere.


\section{Methods} \label{sec:methods}

In this section, we begin with a 2D analytical approximation of the effect of self-heating. We then describe the 3D thermophysical model used to calculate temperatures, the incorporation of 3D shape models/DEMs, and the methods used to couple the model to sublimation, sputtering, the effects of the water exosphere and lag layer formation. Then, we describe the methods used to  produce albedo maps from predicted sublimation rates including the effects of different exosphere densities. 

\subsection{2D Analytical Estimate}

Analytical calculations can provide a high-level estimate of the effect of thermal segregation in double ridge troughs. Local slopes alter the incidence angle of sunlight as well as the radiation received from the surrounding terrain. Applying a methodology adapted from \citet{Aharonson2006SubsurfaceTopography}, we can estimate the flux enhancement for a facet in the trough of a double ridge. The elevation of the Sun above the horizon ($\beta$) can be reproduced from Equation 4 in \cite{Aharonson2006SubsurfaceTopography}, and expressed as $\sin\beta = \cos \Lambda \cos \delta \cos h + \sin \lambda \sin \delta$. Here $\Lambda$ is the latitude, $\delta$ is the solar declination or the angle between the direction of the Sun and the equatorial plane of the body, and $h$ is the hour angle. This can be used to obtain $\beta_{\mathrm{Sun}}$, the angle of the Sun above a sloped surface: 

\begin{equation}
    \sin\beta_{\mathrm{Sun}} = \cos{\alpha} \sin{\beta} - \sin{\alpha}\cos{\beta}\cos{\Delta a}
\end{equation}

where $\alpha$ is the slope angle and $\Delta a$ is the difference between the azimuth of the topographic gradient and that of the Sun. 

The insolation absorbed by the slope can be approximated using the solar constant $S_{0}$, equal to 1361 W/m$^{2}$ (e.g. \cite{Thuillier2003TheMissions, Richard2024AdvancementsContinuity}), the distance from the Sun $R \approx 5.2$ AU, albedo $a$ and $ \beta_{\mathrm{Sun}}$:

\begin{equation}
    Q_{\mathrm{solar}} = \frac{S_{0}}{R^2}  (1 - a) \sin \beta_{\mathrm{Sun}}
\end{equation}

In the case of a trough, the illuminated wall also receives secondary heating from the adjacent wall. Adapting the formulation from \cite{Aharonson2006SubsurfaceTopography} for heating from a surface in the field of view, we can estimate the flux received from the other trough wall: 

\begin{equation}
    Q_{\mathrm{wall}} = \sin^2 \left({\frac{\pi - \theta_{\mathrm{trough}}}{2}}\right) \left( \epsilon\sigma T^{4}_{\mathrm{land}} + (1 - a) Q_{\mathrm{reflected}} \right)
\end{equation}

Here, the first term weights emission according to the incidence angle and integrates over the solid angle subtended by the other surface. $\theta_{\mathrm{trough}}$ is the opening angle of the trough, $\epsilon$ is the emissivity, and $T_{\mathrm{wall}}$ and $Q_{\mathrm{reflected}}$ are the temperature and reflected light from the adjacent wall, respectively. The resulting radiative equilibrium temperature is:

\begin{equation}
    T_\mathrm{eq} = \left[ \frac{Q_\mathrm{solar} + Q_\mathrm{wall}}{\epsilon \sigma} \right]^{1/4}
\end{equation}

We consider the simple case of an equatorial trough with opening angle of 60 degrees at noontime, $\delta = 0$ and uniform albedo of 0.6. In this case, one wall receives up to 7.5 $\mathrm{W~m^{-2}}$ from reemitted IR and $\sim2$ $\mathrm{W~m^{-2}}$ from reflected light. This can produce temperatures  $T_\mathrm{eq} \sim$140 K, about 10 K higher than the equatorial noontime temperature found in \cite{Spencer1999TemperaturesAnomalies} for similar albedo. Heating rates are highly sensitive to the geometry of the trough, the solar incidence angle and the properties of the surface materials, but this calculation indicates self-heating can have a substantial impact on temperature and ice sublimation rates, as well as on properties like the crystallinity of the ice.

\subsection{3D Thermophysical Model}
A model that includes shadowing, topography and surface-to-surface radiation exchange is required to represent the temperature contrasts that can arise from topography \citep{Paige2010DivinerRegion}. We thus adapted the 3D ray tracing thermophysical model from \cite{Sorli2025AFG3}, which includes secondary heating effects. Though initially developed for binary asteroid systems, this model is also applicable to terrain and topography on larger bodies by generating a triangular mesh from digital elevation models. The model calculates temperatures using 1) 1D heat conduction in the subsurface 2) direct insolation at the surface 3) infrared emission 4) visible light reflection 5) scattered IR radiation and 6) geothermal heat flow. These terms are represented in the following surface energy balance equation: 

\begin{align}
    \kappa \frac{\partial T}{\partial z}+(1-A)Q_\mathrm{solar}^\mathrm{direct} - \epsilon \sigma T_\mathrm{s}^4 + (1-A)Q_\mathrm{solar}^\mathrm{diffuse} + \epsilon Q_\mathrm{IR}^\mathrm{diffuse} = 0
\label{eq:heat_bdry}
\end{align}

where $\kappa$ is the thermal conductivity, $\frac{\partial T}{\partial z}$ is the gradient of temperature $T$ with depth $z$, $\epsilon$ is the thermal emissivity, $A$ is the albedo, and $T_{\mathrm{s}}$ is the surface temperature. The first term represents the heat flow. Note that the geothermal heat flux contributes to the temperature gradient and is specified at the bottom boundary. The second term represents absorbed insolation. For the ith facet of a mesh, $Q_\mathrm{solar}^\mathrm{direct}$ is calculated according to Eq. \ref{eq:insolation}, where $L_{\odot}$ is the solar luminosity and $R$ is the distance from the Sun at each timestep. The incidence angle of sunlight is taken into account with the dot product between the solar vector $\hat{S_{\mathrm{i}}}$ and the facet normal $\hat{n_{\mathrm{i}}}$. $V_{\mathrm{i}}$ is the visibility term as determined by ray tracing. It is 0 if the facet is shadowed and 1 if it is illuminated.

\begin{equation}
    Q_\mathrm{solar, i}^\mathrm{direct}= V_{\mathrm{i}} \frac{L_{\odot}}{4\pi R^{2}} \hat{S_{\mathrm{i}}} \cdot \hat{n_{\mathrm{i}}}
    \label{eq:insolation}
\end{equation}

Self-heating is included by the fourth and fifth terms in Equation \ref{eq:heat_bdry}, which represent visible reflected light and scattered IR light, respectively. In this model we assume Lambertian scattering and incorporate surface-to-surface radiation through view factors. The view factor $F_{\mathrm{ij}}$ is the fraction of radiation emitted or scattered from a facet with discrete area $\Psi_{\mathrm{i}}$ that reaches another facet with discrete area $\Psi_{\mathrm{j}}$. The expression for $F_{\mathrm{ij}}$ is shown in Equation \ref{eq:simpleVF} \citep{Rezac2020AccuracyAsteroids}. It includes the product of the cosines of the angles $\phi_{\mathrm{ij}}$ and $\phi_{\mathrm{ji}}$ between the normals of the facets and the distance between two discrete areas, $d_{\mathrm{ji}}$ : 

\begin{equation}
    F_{\mathrm{ij}} = \frac{\Psi_{\mathrm{j}} \cos\phi_{\mathrm{ij}} \cos\phi_{\mathrm{ji}}}{\pi d^{2}_{\mathrm{ji}}}
    \label{eq:simpleVF}
\end{equation}

The above equation is valid when $\sqrt{A_{j}} \gg d_{\mathrm{ji}}$. For the case where this is not true and facets are very close together, a singularity can be reached as $d_{\mathrm{ji}}$ approaches zero. A more robust form is given by the `M2' method of \citet{Rezac2020AccuracyAsteroids}. We utilize the M2 method, shown in Equation \ref{eq:m2VF}, in our model. In the limit of a large $d_{\mathrm{ij}}$, this equation is equivalent to that shown in Equation \ref{eq:simpleVF}.  

\begin{equation}
    F_{\mathrm{ij}} = \frac{4\sqrt{\Psi_{\mathrm{i}} \Psi_{\mathrm{j}} }}{\pi^{2} \Psi_{\mathrm{i}}}\arctan \left[\frac{\sqrt{\pi \Psi_{\mathrm{i}}}\cos\phi_{\mathrm{ij}}}{2 d_{\mathrm{ij}}}\right] \arctan \left[\frac{\sqrt{\pi \Psi_{\mathrm{j}}}\cos\phi_{\mathrm{ji}}}{2 d_{\mathrm{ij}}}\right]
    \label{eq:m2VF}
\end{equation}

We multiply the calculated view factors by the reflected incident flux and temperatures of other facets in the mesh at each timestep of the model to include the effects of self-heating. Thus for the ith facet, $Q_\mathrm{IR, i}^\mathrm{diffuse} = \sum_{\mathrm{j}} F_{\mathrm{ij}}\left(\epsilon \sigma T^{4}_{\mathrm{j}} \right)$ and $Q_\mathrm{solar, i}^\mathrm{diffuse} = \sum_{\mathrm{j}} F_{\mathrm{ij}}Q_{\mathrm{ref, j}}$, where $Q_{\mathrm{ref, j}}$ is the flux reflected off the jth facet. For results reported in this work, all facets use the same albedo $A$, but varied albedo values can be included if known. 

Following \citet{Vasavada2012LunarExperiment} and \citet{Hayne2015ThermalTopography}, we use an exponential density profile. With the appropriate boundary conditions (i.e., Equation \ref{eq:heat_bdry}), we solve the 1D heat equation at each facet and depth grid point using a standard discrete representation of the temporal and spatial derivatives based on an Eulerian (forward-time, centered space) scheme. This approach is accurate for realistic choices of thermophysical properties, including the nonlinear effects of temperature-dependent heat capacity and conductivity. However, it has the drawback (compared to implicit schemes) of conditional numerical stability, which must be determined based on the numerical grid spacing and thermophysical properties.

Heat conduction in the subsurface is treated in 1D. The thermal skin depth, $z_{\mathrm{s}}$, describes how far a thermal wave penetrates into a planetary surface. For a thermal wave with period P:
\begin{equation}
    z_{\mathrm{s}} = \sqrt{\frac{\kappa}{\rho c} \frac{P}{\pi}}
\end{equation}

Here $\rho$ is the density and $c$ is the specific heat capacity. If $z_{\mathrm{s}}$ is larger than the scale of the facets of the DEM, lateral heat conduction between facets must be included to ensure valid results \citep{Rubanenko2017StabilityTopography}. However, the thermal skin depth on the surface of Europa is small, only about 4 cm, which is much less than the resolution of the shape model. A 1D approach to subsurface heat conduction is thus sufficiently accurate for our model.

The thermophysical parameters used for Europa are shown in Table \ref{tab:params}. The calculated timestep to ensure model stability is $\sim$1321 seconds, or just over 22 minutes. We use the coefficient values of 90.0 and 7.49 used to solve the empirical temperature law for heat capacity for ice from \citet{Klinger1981SomeNuclei}. The thermal inertia values reported in the table were selected manually to reproduce the maximum and minimum temperatures from \citet{Spencer1999TemperaturesAnomalies}. With these values, the average thermal inertia between the surface and the skin depth $z_{\mathrm{s}}$, weighted by depth, is about 50.5 $\mathrm{J~m^{-2}~K^{-1}~s^{-1/2}}$. This value falls within the range of thermal inertia values determined by \citet{Spencer1999TemperaturesAnomalies} (45 - 70 $\mathrm{J~m^{-2}~K^{-1}~s^{-1/2}}$) and \citet{Rathbun2010GalileoProperties} (45 - 150 $\mathrm{J~m^{-2}~K^{-1}~s^{-1/2}}$). As these values are lower than for solid ice, they represent an unconsolidated, or fragmented, regolith layer over ice. 

\begin{table}
    \centering
    \begin{tabular}{|p{\dimexpr0.25\linewidth\relax} c c p{\dimexpr0.25\linewidth\relax}|}
    \hline 
    \textbf{Property} & \textbf{Value} & \textbf{Unit (If Applicable)} & \textbf{Source} \\ 
    \hline
    \hline
        Bond Albedo & 0.6 &  & Adapted from \cite{Mergny2025AEuropa} and \cite{dePater1984VLASatellites} \\
        \hline
        Basal Heat Flow & 0.03 & $\mathrm{W / m}^{2}$ & \cite{Ruiz2005TheEuropa}\\
        \hline
        Emissivity & 0.9 & & \cite{Spencer1987SurfacesSpectra}  \\
        \hline
        Specific Heat Capacity (Avg. Surface Temp.) & 900 & $\mathrm{J~kg~K^{-1}}$ & \cite{Klinger1981SomeNuclei}\\
        \hline
        Heat Capacity Polynomial Coefficients &  90.0, 7.49 & & \cite{Klinger1981SomeNuclei} \\
        \hline
        Thermal Inertia at Surface &  13 $\pm$ 50 & $\mathrm{J~m^{-2}~K^{-1}~s^{-1/2}}$ & Values reproducing Galileo PPR data from \cite{Spencer1999TemperaturesAnomalies} \\
        \hline 
        Thermal Inertia at Depth &  63 $\pm$ 40 & $\mathrm{J~m^{-2}~K^{-1}~s^{-1/2}}$ & Values reproducing Galileo PPR data from \cite{Spencer1999TemperaturesAnomalies}\\
        \hline 
        Density at Surface & 100 &  $\mathrm{kg / m^{3}}$& Values reproducing Galileo PPR data from \cite{Spencer1999TemperaturesAnomalies} \\
        \hline 
        Density at Depth & 450 & $\mathrm{kg / m^{3}}$& Values reproducing Galileo PPR data from \cite{Spencer1999TemperaturesAnomalies}\\
        \hline 
    \end{tabular}
    \caption{Thermophysical and compositional parameters used by the model to calculate temperatures and by the lag model to calculate reddening timescales. Here "at depth" refers to a depth $z << H$, where $H$ is the H-parameter related to thermal inertia from \cite{Hayne2017GlobalExperiment}, and which describes the increase in density and conductivity with depth.}
    \label{tab:params}
\end{table}

\subsection{Shape and Digital Elevation Models}
\begin{figure}
    \centering
    \includegraphics[width=0.8\linewidth]{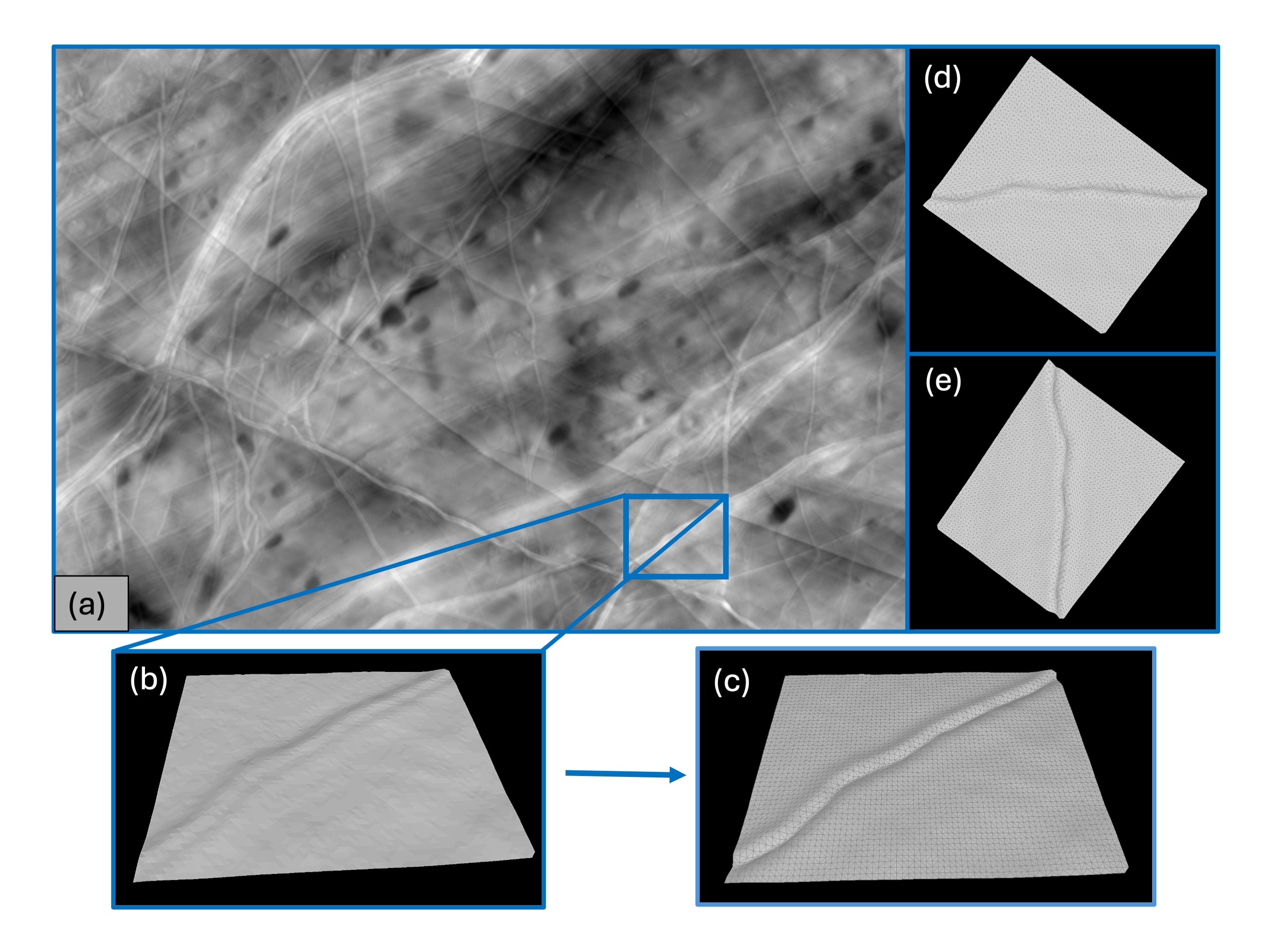}
    \caption{Examples of shape models used or generated in this work. (a) The initial DEM of the Rhadamanthys terrain \citep{Schenk}. (b) The slice removed from the original DEM, which includes a prominent double ridge. Due to resolution constraints, the topography of the ridge is flattened (c) The slice from (b) with a trough manually deepened to approximately the same depth as the ridge height. (d) and (e) show the same shape model rotated so that the trough orientations are roughly east-west and north-south respectively.}
    \label{fig:shapes}
\end{figure}

The 3D thermophysical model requires a shape model in the form of a triangulated mesh. Here, we utilize a digital elevation model (DEM) of Europa’s Rhadamanthys terrain with approximately 1.8 million facets and a resolution of $\sim 100$ meters per pixel \citep{Schenk}. We generate a mesh from the set of cartesian points contained in the DEM through Delaunay triangulation, which ensures the surface is represented as a convex hull with no overhangs or voids. To reduce computational requirements, we extract portions of the larger DEM which contain distinct double ridges, as shown in Figure \ref{fig:shapes}, panel (b). 

Due to limited spatial resolution of the available DEMs, topographical features including double ridges and central troughs are artificially flattened in the original mesh. Previous work suggests that trough heights can range from 10s of meters up to about 300 to 400 meters \citep{Coulter2009, Head1999Europa:Model}. Similarly, though trough morphologies and depths remain uncertain, they are generally thought to be of comparable depth to the ridge height, and possibly deeper depending on the formation mechanism (e.g. \cite{Dameron2018EuropanModels, Dombard2013FlankingEuropa}). Thus, troughs are known to be more prominent than they are represented in the DEM. The depth of the trough can affect the magnitude of secondary heating. To consider this effect, we model illumination on both the original DEM, as well as a suite of DEMs modified so that the trough is approximately as deep as the ridges are tall with respect to their base elevation. A comparison of these trough models is shown in Fig. \ref{fig:shapes} panels (b) and (c). 
To investigate thermal effects at different latitudes, we reposition the DEM described above to represent locations at 0-, 30-, 60- and 90-degrees latitude. This allows for study of how darkening of double ridges changes as latitude increases from equator to the pole. Double ridges also display a range of directions. Temperatures on airless bodies can be strongly affected by orientation, and the efficacy of heating in the trough may be altered by the direction of the ridges. We investigated this effect by reorienting the shape model to represent three trough directions: 1) the original southwest to northeast direction 2) east-west (Fig, \ref{fig:shapes}, panel (d)) and 3) north-south (Fig. \ref{fig:shapes}, panel (e)).  

\subsection{Sublimation Rates}\label{sec:sublimation}
Observable reddening seen in images from Voyager and Galileo could result from the growth of an optically thick dark, non-ice layer. We model this layer and the timescale for its formation using temperatures calculated by the 3D thermophysical model. These temperatures are then used to calculate the sublimation rate of water ice $\dot{E}$ in a vacuum according to the standard formula \citep{Estermann1955}:

\begin{equation}
    \dot{E}(T) = p\mathrm{_{v}}(T)\sqrt{\frac{m}{2 \pi RT}}
    \label{eq:edot}
\end{equation}

The $p\mathrm{_{v}}(T)$ term is the water ice equilibrium vapor pressure at a given temperature $T$, $R$ is the universal gas constant and $m$ is the molecular mass of water. We implicitly assume a sticking coefficient of 1.0. To calculate the vapor pressures, we use the relationship: 
\begin{equation}
\log p\mathrm{_{v}}(T) = \frac{A}{T} + B
\label{eq:satpress}
\end{equation} 
Here the empirical constants are $A$ $= -2663.5 \pm$ 0.8, $B$ $= 12.537 \pm 0.011$, $T$ is in units of Kelvin and $p$ has units Pa \citep{Marti1993A250K}. This relationship yields a sublimation rate upper limit, or the value in a vacuum, in units of kg $\mathrm{m^{-2}}$ $\mathrm{s^{-1}}$. The sublimation rates are calculated at each model timestep. 

We can divide the sublimation rate, $\dot{E}$, by the material density at the surface to obtain an expression for the meters sublimated per second, or $\dot L$: 

\begin{align}
    \dot{L} = \frac{\dot{E}}{\rho}
    \label{eq:Ldot}
\end{align}

The change in layer thickness $\Delta L$ can then be calculated over a time period $\Delta t$ using  $\Delta L = \int_{\Delta t} \dot L  dt$. Here, we integrate $\dot L$ at each timestep over a diurnal cycle ($\Delta{t}_{\mathrm{rot}} = 3.1\times 10^5~\mathrm{s}$), producing a diurnal rate of layer thickness change. This can then be used to calculate $\Delta L_{\mathrm{year}}$ in meters per Earth year. 

In the presence of an exosphere, the sublimation rate is reduced based on the water vapor partial pressure, $p_\mathrm{H_2O}$, using:

\begin{align}
    \dot{E}(p_\mathrm{H_2O}, T) = \left[ p_\mathrm{v}(T) - p_\mathrm{H_2O} \right]\sqrt{\frac{m}{2 \pi RT}}
    \label{eq:edot_exo}
\end{align}

We assume that the sublimation rates of non-ice materials are negligibly small compared to that of H$_2$O.

It is useful to note that the water vapor partial pressures are determined by the vapor pressures of the surrounding ice as well as local sputtering rates. Broadly, water vapor partial pressures are dependent on the vapor pressures and sputtering averaged over a region approximately equal to the molecular jump distance. 

\subsection{Sputtering}
Due to the heavily ionized environment generated by the Jovian magnetosphere, sputtering of volatiles is a significant loss mechanism on Europa’s surface.Previous work has modeled the sputtering rate on Europa. \cite{Cassidy2013MagnetosphericEuropa} calculated the sputtering rate as a result of ion bombardment on the surface of Europa as a function of position, determining a global average sputtering rate of $2 \times 10^{27}$ H$_{2}$O s$^{-1}$. \citet{Plainaki2012TheExosphere} found a global water sputtering rate of $1.7 \times 10^{27}$ H$_{2}$O s$^{-1}$.  Others, such as \citet{Addison2021InfluenceWeathering} and \citet{Breer2019EnergeticEuropa}, used hybrid models of kinetic ions and fluid electrons and investigated the effects of draping magnetic field lines. They, along with later work by \citet{Nordheim2022MagnetosphericSurface}, found large scale regional variations in sputtering, with a marked decrease in ion flux at Europa's upstream apex. 

In this work, we consider ablation induced by temperature-dependent sputtering. \citet{Cassidy2013MagnetosphericEuropa} obtained a global sputtering rate of $2 \times 10^{27}$ H$_{2}$O s$^{-1}$, which corresponds to an erosion rate of about $7 \times 10^{-8}$ meters per year. They proposed the following temperature dependence of the total sputtering yield $Y$: 
\begin{equation}
Y = Y(T = 0)(1 + 220e^{-0.06 \mathrm{keV}/k T}) 
\label{eq:yield}
\end{equation}

Here, $T$ is the surface temperature, and $k$ is the Boltzmann constant. A temperature enhancement factor of $Y = 1$ indicates that the mass equivalent of one $\mathrm{H_{2}O}$ molecule is ejected for every ion that impacts the surface. $Y$ is about 2 at Europa's subsolar point temperature of around 130 K, and about 1 at the night side equator temperature of 80 K \citep{Cassidy2013MagnetosphericEuropa}. Equation \ref{eq:yield} applies to the total sputtering yield, including O$_{2}$ and H$_{2}$ that can result in net ablation from the surface, as well as $\mathrm{H_{2}O}$. 

For each temperature output by the model, we calculate the associated sputtering yield $Y$ and multiply by the global rate to obtain sputtering rates. This allows for an estimate of sputtering and how it varies through a diurnal cycle for each facet. As with sublimation, this can be used to obtain a sputtering rate in meters per Earth year. Previous work has investigated the regional variations in sputtering (e.g. \cite{Nordheim2022MagnetosphericSurface}), as well as the dependence of sputtering yield on angle of implantation \citep{Vidal2005AngularBombardment}. A detailed treatment of sputtering is beyond the scope of this paper, but will be investigated in future work. 

\subsection{Exchange with the Water Exosphere}
Europa has a thin but significant water exosphere due to sublimation and sputtering of the icy surface materials. As a result, it is primarily composed of the dissociative products O$_{2}$ and H$_{2}$, with a tenuous water component \citep{Plainaki2012TheExosphere,Saur1998InteractionAtmosphere}. H$_{2}$ molecules are sufficiently light that they escape Europa's gravity \citep{Nenon2019EvidenceMeasurements}. In contrast, H$_{2}$O molecules sputtered from the surface usually freeze and are deposited back onto the surface quickly \citep{Smyth2006EuropasImplications, Eviatar1985EuropanPhenomena}. Due to the constant bombardment of the surface, there is active replenishment of the water exosphere. The density of this exosphere should be affected by temperature, as well as by endogenic sources like plumes that can lead to transient increases in material (e.g. those observed by \cite{Roth2014TransientPole} and \cite{Paganini2020AEuropa}). Previous work has also suggested that water ice may freeze out on Europa's night side, leading to a day side water exosphere \citep{Teolis2017PlumePredictions}.

As discussed in section \ref{sec:sublimation}, water vapor partial pressure are critical for modeling sublimation/deposition rates, and yet measurements of Europa's water exosphere are limited. Some, such as those from \cite{Paganini2020AEuropa}, provide density estimates but spatial and temporal variability (e.g., due to plume activity) cannot be easily separated from the background exosphere. Several studies have modeled the exospheric water column density. \citet{Smyth2006EuropasImplications} and \citet{Shematovich2005Surface-boundedEuropa} both apply a 1D Monte Carlo multi-species approach and include the effects sublimation and scattering sources. \citet{Shematovich2005Surface-boundedEuropa} includes multiple source rates, thus producing a range of density values. \citet{Teolis2017PlumePredictions} applied a Monte Carlo exosphere model that considers sputtered, radiolytic and plume sources, as well as notable effects like surface adsorption and re-sputtering of adsorbed materials. Lastly, \citet{Vorburger2018EuropasContribution} modeled the Europan exosphere from first principles and without fitting to observations, but with inclusion of sputtered and sublimated particles that follow a Maxwellian thermal energy distribution. 
 
 Model exosphere densities reported in the literature, which we reproduce in Table \ref{tab:exovals}, generally fall within two groups. One represents higher modeled column densities of $10^{18} – 10^{19}$ m$^{-2}$ \citep{Teolis2017PlumePredictions, Teolis2017WaterBodies,Plainaki2012TheExosphere}. Others are distinctly lower, around $10^{16}$ m$^{-2}$ \citep{Vorburger2018EuropasContribution,Shematovich2005Surface-boundedEuropa,Smyth2006EuropasImplications}. Here, we investigate a range of modeled exosphere column density values covering both of these groups. We consider the lower bound of the estimate from \citet{Shematovich2005Surface-boundedEuropa} as the most tenuous exosphere case, as well as the mid-range values from \citet{Smyth2006EuropasImplications} and \citet{Vorburger2018EuropasContribution}. Lastly, we also consider the global average of the day and night side water column density value from \citet{Teolis2017PlumePredictions}, which provides the upper bound of our modeled water exosphere column densities.
 
Herein, we only consider deposition of water molecules as most dissociative  H$_{2}$ escapes, and the O$_{2}$ which remains does not stick efficiently to the surface \citep{Eviatar1985EuropanPhenomena}.
 
To model deposition from the exosphere, we convert exospheric column densities to partial pressures using $p_\mathrm{H_2O} = M g$, where $M$ is the column density (kg m$^{-2}$) and $g$ is Europa's surface gravity. In contrast to a vacuum environment, higher exospheric partial pressure of water will raise the temperature at which sublimation dominates over deposition. This in turn will decrease the temperature at which water condenses onto the surface as ice, making sublimation less effective as an ablation mechanism. In locations where the mean saturation vapor pressure is lower than the mean water vapor partial pressure, i.e., $\overline{p_\mathrm{v}(T)} < \overline{p_\mathrm{H_2O}}$, net deposition occurs rather than sublimation. Here, we calculate the pressure of the exosphere at the surface by calculating the temperature dependent vapor pressure from Eq. \ref{eq:satpress} and subtracting the partial pressure from the water column densities to calculate the sublimation rate (Eq. \ref{eq:edot_exo}) using that pressure for each model timestep. For some higher values of the exosphere column density, net deposition is widespread, and conversely for lower exosphere water vapor values, sublimation dominates.

\begin{table}
    \centering
    \begin{tabular}{|cc|}
    \hline 
    \textbf{Reference} & \textbf{Column Density (m$^{-2}$)}\\ 
    \hline
    \cite{Shematovich2005Surface-boundedEuropa} & 9.2 -- 87 $\times 10^{15}$\\
    \hline
    \cite{Smyth2006EuropasImplications} & 2.30$\times 10^{16}$ \\
    \hline
    \cite{Vorburger2018EuropasContribution} & 6.02$\times 10^{16}$ \\
    \hline
    \cite{Teolis2017PlumePredictions} &  \\
    Day & 7.94$\times 10^{18}$ \\
    Night & 6.3$\times 10^{16}$ \\ 
    Average & 4.0$\times 10^{18}$ \\
    \hline 

\end{tabular}
    \caption{Modeled water exosphere column density values in m$^{-2}$}
    \label{tab:exovals}
\end{table}

\subsection{Low Albedo Lag Formation}\label{sec:albedo}
The observable reddening observed in images of Europa’s surface acquired by Galileo and Voyager imply the existence of a non-ice layer composed of salt hydrate minerals \citep{McCord1999HydratedInvestigation} and/or sulfuric acid hydrates \citep{Carlson2005DistributionHydrate}, which are concentrated in patterns on global and local scales \citep{McCord2010HydratedInvestigation}. Our objective is to test whether such a layer could be formed through sublimation as a ``lag" layer where the non-ice constituent is concentrated relative to the original ice with impurities. To calculate the timescales for lag layer formation, we use the modeled sublimation and sputtering rates as discussed in the previous sections. Ablation rates due to each process can be added to determine a net rate of ablation or deposition over one Earth year, $\dot{L}_{\mathrm{year}}$.

\begin{figure}
    \centering
    \includegraphics[width=0.9\linewidth]{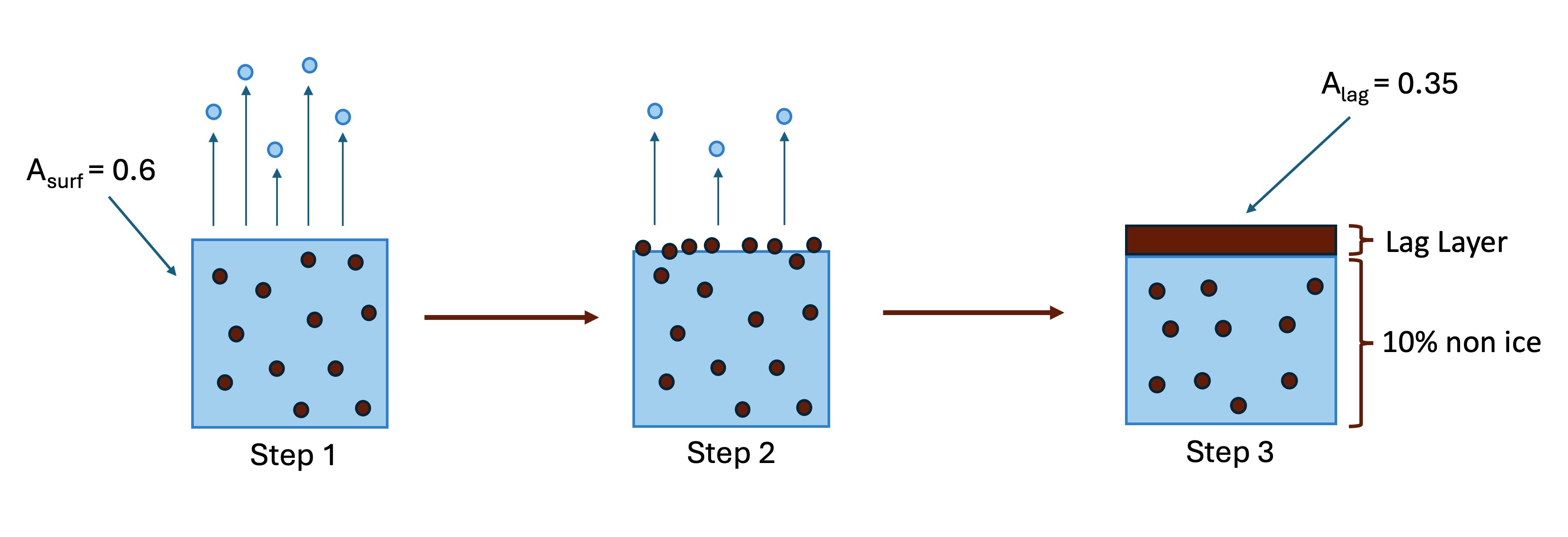}
    \caption{A diagram representing how the lag formation process occurs within the model. Ice properties are modeled in bulk in this formulation, while the non-ice material are modeled as distinct components with an assumed particle size of $1~\mu$m. In Step 1, we begin with a mixture of non-ice material and water ice with a bulk bond albedo of 0.6. The dark material comprise 10\% of the surface material, with the rest being water ice. This water ice sublimates or sputters according to temperatures and rates calculated by the model. This preferentially leaves behind dark material as water ablates through sublimation or sputtering (Step 2). After sufficient loss of water, the remaining dark material layer grows in thickness to form a dark lag layer (Step 3). Once reaching an optically thick depth, this layer appears dark, with an albedo of 0.35.}
    \label{fig:lagdiagram}
\end{figure}

For a lag layer to be optically thick, it must reach a critical depth $L_{\mathrm{crit}}$ where the optical depth $\tau \sim 1$. The process for lag formation used in this work is illustrated in the diagram shown in Figure \ref{fig:lagdiagram}. We assume a homogeneous mixture initially containing 10\% non-ice material represented as dust grains approximately 1 $ \mathrm{\mu m}$ in size \citep{Tomlinson2022CompositionEuropa}. This is represented by concentration $c = 0.1$. Roughly 10 $\mu \mathrm{m}$ of sublimation is thus required to obtain an optically thick 1 $\mu \mathrm{m}$ lag layer. The timescale for formation of this lag layer can thus be determined using $c$ and $\dot{L}$ for each facet in the mesh. 

Notably, the dust grain size is assumed for the purposes of this study. No grain sizes are assumed for water ice, which is instead modeled in bulk. In reality, larger grains may be present on Europa \citep{Mishra2021AEuropa, Ligier2016VLT/SINFONICOMPOSITION}. Different non-ice grain sizes would affect the number of particles in an optically thick layer. Lag thickness will scale roughly linearly with grain size, so that a factor of 2 increase in grain size will lead to $\sim$ 2x increase in lag layer thickness.  

\subsection{Radiative Transfer Model for Surface Albedo}
For dark material deposits to be detectable in the images, thermal segregation must produce an optically thick lag layer dominated by the lower albedo component. The exact composition of the dark component is not important for our analysis, so long as this component is more absorbing than water ice. We thus estimate albedo changes corresponding to lag thicknesses generated by the thermophysical model, and use these results to produce albedo maps.

The radiative transfer model calculates the directional hemispheric albedo using the $\delta$-Eddington two stream approximation for a two-layer medium \citep{Joseph1976TheTransfer, Wiscombe1980ASnow}. Here, we assume the bottom layer is composed of ice mixed with 10\% non-ice material with optical properties required to yield a Bond albedo of 0.6 for the mixture. The top layer, assumed to be composed of discrete grains with optical properties that can be modeled with Mie theory, is composed of pure dark material with a thickness produced by the thermal model. In this layer, we assume the dark material is made up of spherical dust grains $\sim 1~\mathrm{\mu m}$ in radius.

The optical depth $\tau$ of the top lag layer can be written as the following:

\begin{equation}
    \tau = n \sigma_{\mathrm{ext}} L
\end{equation}

Here $n$ is the number density of the low albedo material in particles per cubic meter, $\sigma_{\mathrm{ext}}$ is the extinction cross section, and $L$ is the thickness of the layer produced by the model over a given timescale, i.e., $L = c \dot{L}\Delta t$, with $\dot{L}$ given by Equation \ref{eq:Ldot} and the concentration $c = 0.1$, or 10\% non-ice. Assuming the dust particles are spherical with a radius $r_{\mathrm{eff}} =$ 0.5 $\mu$m, the number density $n$ can also be written as:

\begin{align}
    n &\approx \frac{\left( 1 - \mathrm{porosity} \right)}{\left(\mathrm{volume~of~particle} \right)}\\
    &= \frac{3\left(1 - \phi \right)}{4 \pi r_{\mathrm{eff}}^{3}}
\end{align}

Simple cubic packing of monodisperse spherical grains corresponds to porosity $\phi \approx 0.5$. The extinction cross section can also be written in terms of the extinction efficiency $Q_{\mathrm{ext}}$, such that: 

\begin{equation}
    \sigma_{\mathrm{ext}} =Q_{\mathrm{ext}} \pi r_{\mathrm{eff}}^2
\end{equation}

The optical depth of the top layer can then be written as:
\begin{equation}
    \tau = \frac{3}{4}\frac{Q_{\mathrm{ext}}}{r_{\mathrm{eff}}} \left(1-\phi \right) L
\end{equation}
The upper layer optical depth is calculated for each facet using the lag layer thickness in meters produced over a given number of years. The facet-by-facet optical depths are then applied to calculate the surface albedo using the $\delta$-Eddington model. Additional parameters for this model include the cosine of the emission angle $\mu$ of the observation, the single scattering albedo  $\varpi_{0}$, the asymmetry parameter $g = \langle cos(\mathrm{\theta_{phase})\rangle}$, and the albedo of the underlying surface $a_{\mathrm{surf}}$. In the limit of geometric optics at a given wavelength $\lambda$ where $x = 2\pi r_\mathrm{eff}/\lambda \gg 1$, the extinction efficiency, single scattering albedo and asymmetry parameter approach constant values. As we are working in this regime ($x \approx 10$), we assume $Q_{\mathrm{ext}} \approx 2$.

For the dust/non-ice component, we use a single scattering albedo $\varpi_{0} = 0.88$. This matches the deduced single scattering albedo in the Voyager clear (0.47 $\mathrm{\mu m}$ filter) for the lowest albedo mottled terrains \citep{Domingue1992Disk-resolvedTerrains}. We also use an asymmetry parameter $g = 0.2$ (i.e., mostly isotropic scattering). Using the radiative transfer model described above, these values produce an albedo of $\sim$ 0.35 in the limit that the lag layer reaches a thickness where the optical depth $\tau \gg 1$. This value is utilized as it is comparable to that found for the darkest regions on Europa such as the trailing hemisphere chaos and lenticulae \citep{Mergny2025AEuropa, Helfenstein1998GalileoEffect}. Finally, with these parameter choices and lag layer optical depths, the model produces a directional hemispheric albedo for each facet, which we used to produced albedo maps for discrete time intervals. 

It should be noted that non-ice material on Europa may have varied and larger particle sizes. In the limit of geometric optics considered here, at an optical depth of $\tau = 1$ the layer thickness scales roughly as $L \sim r_{\mathrm{eff}}$. Thus, a change in the particle size should correspond with an approximately linear difference in layer thickness to achieve optically thick lag.


\section{Results}

\subsection{Surface Temperatures}

\begin{figure}[H]
    \centering
    \includegraphics[width=0.9\linewidth]{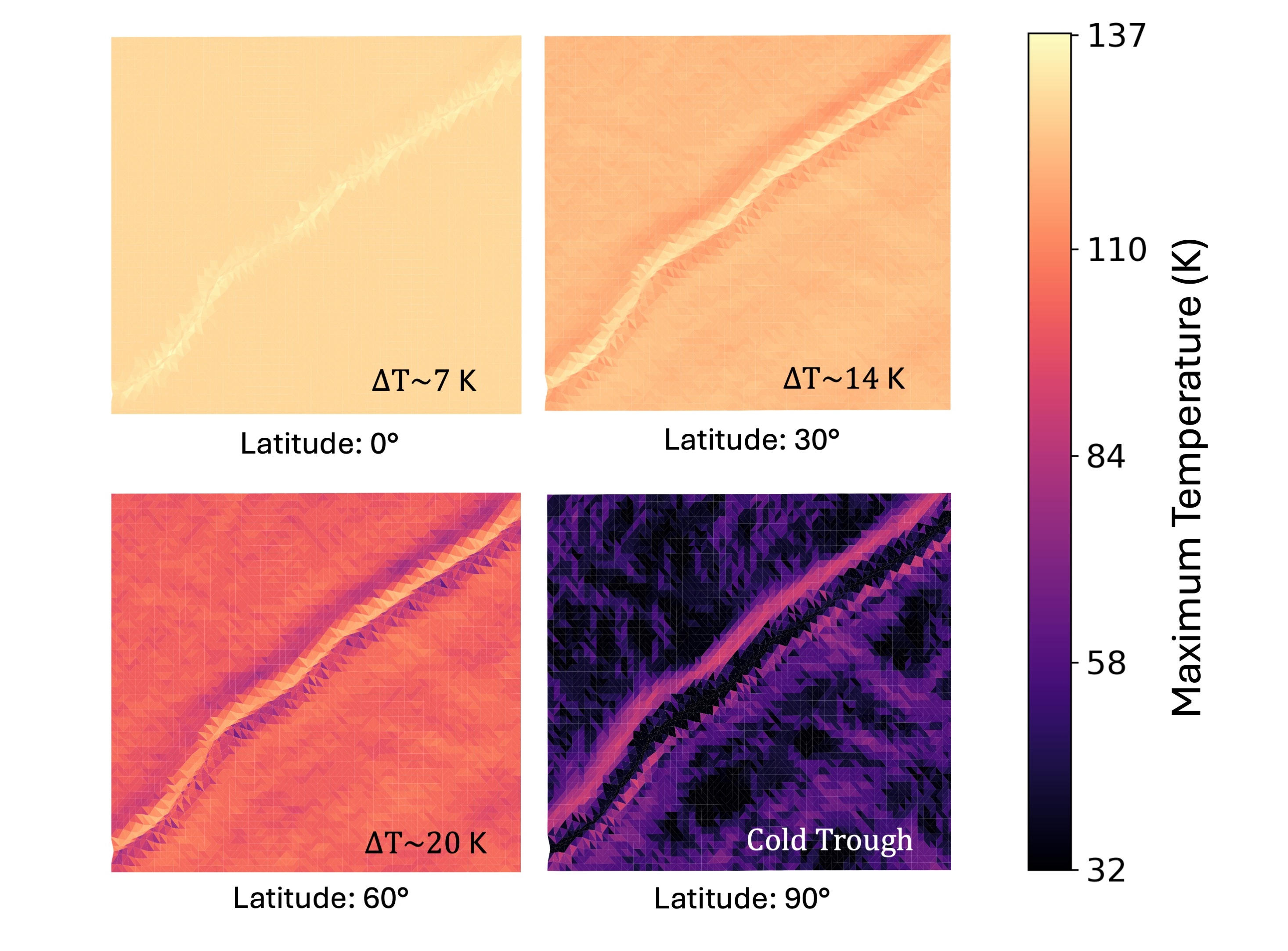}
    \caption{Maximum modeled temperatures for each facet in the manually-deepened trough with varying latitude, using the original trough orientation. Background temperatures agree well with Galileo PPR temperatures. Notably, the troughs exhibit higher temperatures than their surroundings as a result of self-heating. This is not true of the pole, where topography of the ridges causes shadowing and comparably lower temperatures. The approximate difference in maximum temperatures inside and outside of the trough is included in the bottom right of each image.}
    \label{fig:maxtemps}
\end{figure}

Sublimation and sputtering are both temperature dependent processes, making thermophysical modeling critical to understanding their relative contributions to the mass balance of Europa's surface ice. We modeled double ridge terrain at 0$^\circ$, 30$^\circ$ 60$^\circ$ and 90$^\circ$ latitude using the thermophysical properties detailed in Table \ref{tab:params}. Fig. \ref{fig:maxtemps} shows maps of the modeled maximum temperatures on a prominent ridge positioned at each of the four latitudes. Note that this figure shows results with the manually-deepened trough. Modeled background temperatures for areas outside the trough reach $\sim$130 K at the equator. At higher latitudes, background temperatures are lower, with maximum background temperatures of $\sim$122 K at $30^{\circ}$,  $\sim$102 K at $60^{\circ}$ and $ < 70$ K at $90^{\circ}$. The model thus agrees well with the daytime surface temperatures measured on Europa by the Galileo PPR instrument, which recorded a maximum equatorial temperature of about 132 K \citep{Spencer1999TemperaturesAnomalies}.

In this section we present the results of the thermophysical model, the lag formation rates as a result of sublimation, sputtering and deposition from the water exosphere, and albedo maps using lag formation estimates. We also discuss what modeled water exosphere densities produce net ablation versus deposition on the surface. 

\begin{table}[ht]
\centering
\renewcommand{\arraystretch}{1.3}

\begin{tabular}{lcccc}
\toprule
\textbf{Latitude} & \multicolumn{3}{c}{\textbf{Absolute $T_{\mathrm{max}}$}}  & \textbf{Background median $T_{\mathrm{max}}$}\\
\cmidrule(lr){2-4}
\cmidrule(l){5-5}
 & Original & North-South & East-West \\
\midrule
$0^{\circ}$ & 137.0 K & 137.1 K & 137.2 K & $\sim129$ K\\
$30^{\circ}$ & 135.9 K & 134.2 K & 137.8 K & $\sim122$ (NS) - 123 K (Orig., EW) \\
$60^{\circ}$ & 123.6 K & 121.7 K & 128.7 K & $\sim$ 102 (Orig.) - 105 K (NS, EW) \\
$90^{\circ}$ & 92.9$^{*}$ K &  $92.8^{*}$ K & $92.8^{*}$ K  & $\sim$45 K\\
\bottomrule
\end{tabular}
\vspace{0.5\baselineskip}
\caption{Maximum temperatures observed in the trough and median maximum background temperatures at a range of latitudes for the original southwest-to-northeast orientation, the north-south orientation and the east-west orientation. Temperatures listed are the maximum experienced by a facet in the trough over the course of one diurnal cycle. The  median maximum background temperatures are given on the right. If the values are not the same for all trough orientations, they are specified for original (Orig.), north-south (NS) and east-west (EW). $^{*}$At the poles, the trough remains shadowed. Thus, the peak trough temperatures occur at the top of the trough along the crest of the ridge, where light hits. Temperatures in the bottom of the trough remain beneath $\sim$45 K. }
\label{tab:maxTroughTemps}
\end{table}

Sublimation rate is particularly sensitive to the peak temperatures experienced by a surface, making maximum diurnal temperatures a useful metric for estimating mass balance. We find that for all non-polar latitudes, the troughs exhibit higher maximum temperatures than those of the background terrain. This is shown in Fig. \ref{fig:maxtemps}, which uses the deeper trough. As discussed previously, these locally elevated temperatures in the ridge troughs and other topographic depressions are the result of self-heating, with reflection and re-emission more substantial than on the surrounding terrain. We find that the effect is minimal for the original unaltered low-resolution shape model, which has few topographical concavities. In this case, the trough is less than 5 K warmer than the surroundings, and the highest temperatures are found on the flanks of the double ridge. Self-heating is, however, significantly stronger for the trough with comparable depth to ridge height. Thus, all results reported hereafter refer to the shape model with increased (more realistic) trough depth. At the equator, the original southwest-northeast trough reaches peak temperatures of up to about 137 K. At $30^\circ$, peak temperatures are slightly lower, with maximum temperatures in the range $\sim 129 - 135$ K. By $60^\circ$, peak temperatures in the trough are $\sim110 - 124$ K. In each panel of Figure \ref{fig:maxtemps} we provide the temperature difference $\Delta T$ relative to the local background. $\Delta T$ is $\sim 7$ K at the equator, and this temperature anomaly increases with increasing latitude. At  $30^\circ$, $\Delta T$ is about 14 K, while at $60^\circ$ $\Delta T$ increases to $\sim 20$ K. This indicates that self-heating may be a powerful mechanism for causing locally warmer temperatures and enhanced sublimation in the middle and higher latitudes. 

Europa's extreme polar regions are an exception to this trend. At the highest latitudes, topography can cast perpetual shadows \citep{Paige2010DivinerRegion}. We find that the interiors of ridge troughs when in perpetual shadow are usually less than 50 K, with the warmest temperatures found on the equator-facing outer flanks of the ridges.  Thus the interiors of ridges at Europa's poles may be too cold to experience H$_2$O sublimation, which is only significant above $\sim 110$~K. These results point to how topography may also lead to locally lower temperatures, producing cold trap areas where volatiles like $\mathrm{CO}_{2}$ may be stable \citep{Zhang2009Cold-trappedOrigins}.

It is worth noting that at the pole, each ridge flank receives direct insolation daily. However, the geometry and steepness of the flank affects their resulting peak temperature. In Figure \ref{fig:maxtemps}, the flank of the bottom ridge is less steep than the flank of the top ridge, leading to comparably lower peak temperatures.

Orientations of topographic features like ridges can affect illumination and heating rates, and thus sublimation and deposition. Accordingly, we also modeled the temperatures of double ridges with varied orientation, including north-south and east-west running troughs. We find that orientation has a small but measurable effect on the maximum temperatures reached in the trough, the results for which are summarized in Table \ref{tab:maxTroughTemps}. Though background temperatures are comparable, the north-south trough is $\sim$ 1 K warmer at the equator when compared to other orientations. However, as latitude increases, the north-south trough runs colder than the other orientations by about 1 - 5 K in the trough. In contrast, the east-west orientation produces the warmest trough at latitudes higher than the equator. The east-west trough has the highest trough temperatures at $30^\circ$ latitude, with a maximum of $\sim$138 K. Critically, the east-west orientation also results in warmer facets in the $60^\circ$ latitude trough than the other orientations. The maximum temperature recorded in the east-west $60^\circ$ trough is $\sim$129 K, about 5 K higher than that of the original trough orientation at the same latitude. This, in turn, will lead to higher sublimation rates in the east-west trough than the other orientations at higher latitudes. By running east to west, this orientation allows for longer durations where the Sun fully illuminates the trough, which also leads to larger amounts of self-heating.

\subsection{Sublimation and Sputtering}

\begin{figure}
    \centering
    \includegraphics[width=1.0\linewidth]{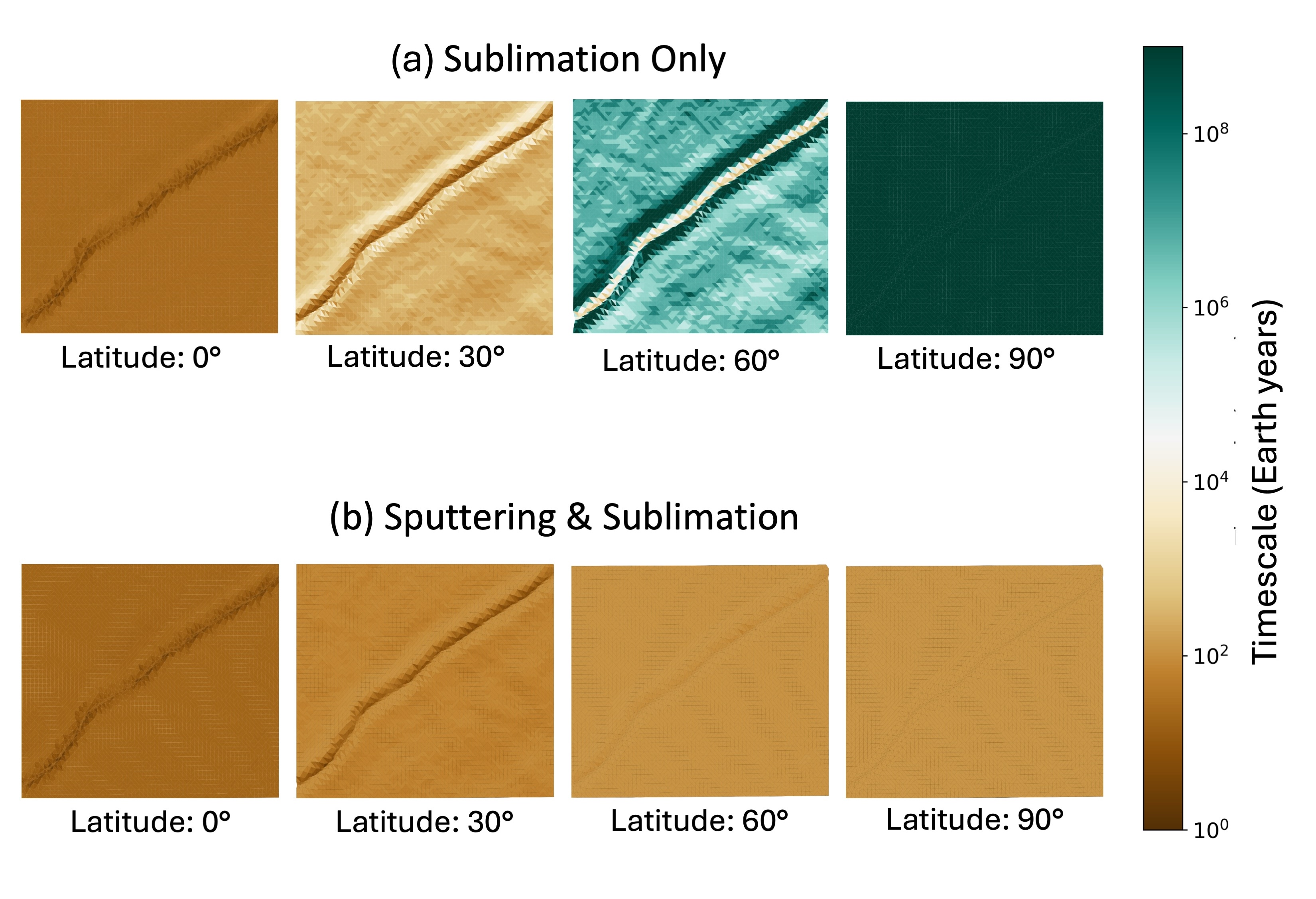}
    \caption{Timescales in Earth years to produce a 1 $\mu$m optically thick lag layer for each facet in a vacuum. The top row shows the timescales for each modeled latitude using only sublimation as an ablation mechanism. The bottom row shows lag formation timescales for sublimation and sputtering combined. Latitude increases from the equator in the far left panels to the pole in the far right panels. From the equator to the middle latitudes, sublimation dominates. Beyond the middle latitudes, sputtering dominates, resulting in a nearly flat rate at the pole.}
    \label{fig:timescales}
\end{figure}

Higher temperatures in the trough produce higher sublimation rates, resulting in shorter timescales for ice in  the trough to sublimate and form an optically thick lag layer than its surroundings. We find that lag formation occurs more quickly inside and on the walls of the troughs for latitudes ranging from $0^\circ$ to $60^\circ$. Figure \ref{fig:timescales} shows the timescales for formation of a 1 $\mu\mathrm{m}$ thick lag layer in a vacuum, neglecting the exosphere. Results for models including sublimation only are shown for comparison with those including both sublimation and sputtering, for each of the four latitudes considered. 

We find that sublimation is the dominate mechanism for H$_2$O ablation within troughs from the equator to the mid-latitudes. In a hypothetical vacuum, this would lead to rapid lag formation, with the warmest facets in the trough acquiring a 1 $\mathrm{\mu}$m thick layer due to sublimation in less than 10 years, with most of the trough forming such a lag layer in a few tens of years. At lower latitudes, sublimation rates on the background terrain would be rapid as well, forming a lag layer in about $\sim10^{1}$-$10^{3}$ years. However, even under vacuum, sublimation rates would be negligible at latitudes higher than about $60^\circ$. At $60^\circ$, sublimation is a significant effect only in the trough, which is warmer by $\sim$ 10 - 20 K than the background terrain. Sublimation rates in the trough are between $10^{-10}$ - $10^{-8}$ kg~m$^{-2}$ per diurnal cycle, while the background terrain has a median sublimation rates of $10^{-12}$ kg~m$^{-2}$ per diurnal cycle. This leads to lag formation timescales of $\sim10^{2}$-$10^{4}$ years. Outside of the trough, sublimation in isolation produces a lag layer in $\sim10^{5}$-$10^{8}$ years.

From $60^\circ$ to the poles, sputtering is the dominant mechanism. Though sublimation and sputtering are additive effects in the trough at $60^\circ$,  sublimation is negligible for level terrain and at higher latitudes. Thus the contribution of sputtering reduces lag formation timescales from $> 10^{9}$ years, or effectively stable ice, to $\sim10^{2}$ years at 60$^\circ$ latitude.

These modeling results suggest that thermal segregation could produce detectable albedo changes in troughs on geologically short timescales, with potential for a latitudinal variation of reddening of topographic features on Europa's surface generally. If the initial concentration of non-ice material were to change, the rate of lag formation would scale linearly with the concentration times the fractional area of the particles on the surface. If the concentration reaches very high levels, water particles may no longer be able to escape, and this process would slow.

\subsection{Effects of the Water Exosphere}

Though Fig. \ref{fig:timescales} shows sublimation timescales under vacuum, the presence of an exosphere is expected to reduce the efficiency of sublimation and sputtering in promoting thermal segregation and lag layer formation. We include the effects of a water exosphere using a range of modeled water exosphere densities. Using modeled lag formation rates, we calculate the change in material $\Delta L$ on each facet as a result of sublimation, sputtering and deposition from the water exosphere. 

We find that whether net ablation and resulting lag formation occurs is highly sensitive to the exosphere density value used (cf. Equation \ref{eq:edot_exo}). The highest modeled exosphere density considered, $\sim 10^{18}~\mathrm{H_2O~m^{-2}}$ \citep{Teolis2017PlumePredictions}, produces no ablation and instead results in deposition of H$_2$O equivalent to $\sim 10^{-5}~\mathrm{m}$ of ice per year, including the effects of both sublimation and sputtering. However, it is important to note that less material is deposited in the ridge trough, although net deposition is still observed.

\begin{figure}
    \centering
    \includegraphics[width=0.99\linewidth]{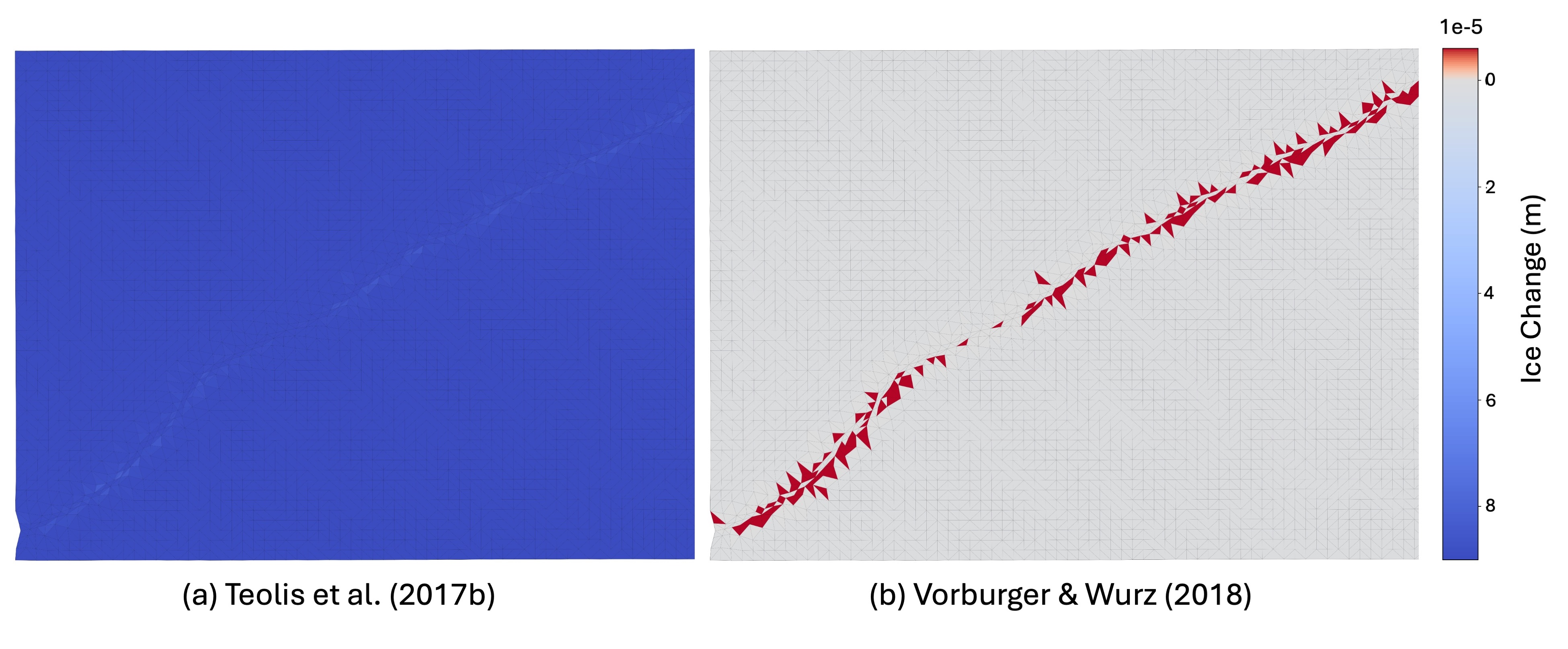} 
    \caption{Expected surface mass balance change over one year using lag thicknesses produced by the thermal model, and including the effects of sublimation, sputtering and the water exosphere. In (a), the exosphere density applied is the average of the day-night value from \citet{Teolis2017PlumePredictions}. This results in net deposition, shown in blue, of order $10^{-5}$ m. The trough experiences slightly less deposition. (b) applies the exospheric density from \citet{Vorburger2018EuropasContribution}, resulting in net ablation in the trough (shown in red) and slight deposition in the surroundings (light gray). For these same plots, normalized to their own range of values and thus showing increased detail, please see Supplementary Materials figures S3 and S4.}
    \label{fig:deposition}
\end{figure}

In contrast, applying the lower water exospheric density estimates of $\sim 10^{16}~\mathrm{H_2O~m^{-2}}$ \citep{Vorburger2018EuropasContribution} results in net ablation throughout the ridge troughs. The warmest facets produce a dark lag layer several microns thick, although most sublimate less. The background terrain experiences net deposition of a thin layer of water ice an order of magnitude thinner than that predicted with exosphere density values from the upper end of the range \citep{Teolis2017PlumePredictions}. A comparison of these results is shown in Figure \ref{fig:deposition}, where blue facet color indicates deposition of ice and red facet color indicates ablation. For the sake of comparison, both plots are shown with the same scale. 

Table \ref{tab:ablation} records whether net ablation occurs at each latitude for each of the modeled exosphere column density values. We find that all modeled densities except for the highest value from \citet{Teolis2017PlumePredictions} result in some amount of ablation at both equatorial and $30^\circ$ latitudes for all trough orientations. The lowest tested value, the lower bound of the range from \citet{Shematovich2005Surface-boundedEuropa}, results in nearly all of the modeled DEM experiencing net ablation and lag formation at the equator. \citet{Smyth2006EuropasImplications} produces heavy ablation in the trough and surroundings, and \citet{Vorburger2018EuropasContribution} leads to ablation tightly aligned with the trough. At $30^\circ$, ablation is observed to some degree in the trough for all trough orientations. At $60^\circ$, some ablation is observed in the trough only for the east-west trough orientation, using the exosphere from \citet{Shematovich2005Surface-boundedEuropa}. A single trough facet experiences a small ($\sim10^{-9}$ m) amount of ablation using the exospheric density from \cite{Smyth2006EuropasImplications}. An exosphere using \citet{Teolis2017PlumePredictions} produces no net ablation at any latitude. No modeled water exosphere density results in net ablation at latitudes higher than $60^\circ$.

\begin{table*}[p]
\centering
\renewcommand{\arraystretch}{1.2}

\begin{subtable}{0.47\textwidth}
\centering
\caption{\cite{Shematovich2005Surface-boundedEuropa}: 9.2e15 m$^{-2}$}
\begin{tabular}{l | c c c c}
\toprule
 & $0^\circ$ & $30^\circ$ & $60^\circ$ & $90^\circ$ \\
\midrule
& \textbf{\underline{Orig.}} \\
Net Abl.                & A & T & N & N \\
$\Delta \mathrm{Ice_{max}}$ & -7.3e-6 & -5.1e-6 & 5.1e-8 & 1.5e-7 \\
$\Delta \mathrm{Ice_{med.}}$ & -7.3e-6 & 7.9e-8 & 1.4e-7 & 2.4e-7 \\[0.6em]

& \textbf{\underline{NS}} \\
Net Abl.                & A & T & N & N \\
$\Delta \mathrm{Ice_{max}}$ & -7.6e-6 & -2.8e-6 & 9.2e-8 & 1.5e-7 \\
$\Delta \mathrm{Ice_{med.}}$ & -7.6e-6 & 4.9e-8 & 1.4e-6 & 2.5e-7 \\[0.6em]

& \textbf{\underline{EW}} \\
Net Abl.                & A & T & T & N \\
$\Delta \mathrm{Ice_{max}}$ & -7.9e-6 & -9.7e-6 & -3.2e-7 & 1.5e-7 \\
$\Delta \mathrm{Ice_{med.}}$ & -7.9e-6 & 5.3e-8 & 1.4e-7 & 2.4e-7 \\
\bottomrule
\end{tabular}
\end{subtable}
\hfill
\begin{subtable}{0.47\textwidth}
\centering

\caption{\cite{Smyth2006EuropasImplications}: 2.3e16 m$^{-2}$}
\begin{tabular}{|c c c c}
\toprule
 $0^\circ$ & $30^\circ$ & $60^\circ$ & $90^\circ$ \\
\midrule
\textbf{\underline{Orig.}} \\
T,S & T & N & N \\
-7.0e-6 & -4.8e-6 & 3.7e-7 & 5.1e-7 \\
-7.0e-6 & 4.0e-7 & 4.9e-7 & 7.2e-7 \\[0.6em]

\textbf{\underline{NS}} \\
T,S & T & N & N \\
-7.6e-6 & -2.8e-6 & 9.2e-8 & 1.5e-7 \\
-7.5e-6 & 4.9e-8 & 1.4e-7 & 2.5e-7 \\[0.6em]

\textbf{\underline{EW}} \\
T,S & T & T & N \\
-7.5e-6 & -9.4e-6 & -4.3e-9 & 5.1e-7 \\
-7.4e-6 & 3.7e-7 & 4.8e-7  & 7.4e-7 \\
\bottomrule
\end{tabular}
\end{subtable}

\vspace{1em}

\begin{subtable}{0.47\textwidth}
\centering
\caption{\cite{Vorburger2018EuropasContribution}: 6.02e16 m$^{-2}$}
\begin{tabular}{l | c c c c}
\toprule
 & $0^\circ$ & $30^\circ$ & $60^\circ$ & $90^\circ$ \\
\midrule
& \textbf{\underline{Orig.}} \\
Net Abl.                & T & T & N & N \\
$\Delta \mathrm{Ice_{max}}$ & -6.1e-6 & -4.0e-6 & 1.2e-6 & 1.5e-6 \\
$\Delta \mathrm{Ice_{med.}}$ & 8.5e-7 & 1.3e-6 & 1.4e-6 & 2.0e-6 \\[0.6em]

& \textbf{\underline{NS}} \\
Net Abl.                & T & T & N & N \\
$\Delta \mathrm{Ice_{max}}$ & -6.5e-6 & -1.6e-6 & 1.3e-6 & 1.5e-6 \\
$\Delta \mathrm{Ice_{med.}}$ & 8.5e-7 & 1.2e-6 & 1.4e-6 & 2.1e-6 \\[0.6em]

& \textbf{\underline{EW}} \\
Net Abl.                & T & T & N & N \\
$\Delta \mathrm{Ice_{max}}$ & -6.7e-6 & -8.5e-6 & 8.4e-7 & 1.5e-6 \\
$\Delta \mathrm{Ice_{med.}}$ & 8.5e-7 & 1.2e-6 & 1.4e-6 & 2.1e-6 \\
\bottomrule
\end{tabular}
\end{subtable}
\hfill
\begin{subtable}{0.47\textwidth}
\centering
\caption{\cite{Teolis2017PlumePredictions}: 4.0e18 m$^{-2}$}
\begin{tabular}{|c c c c}
\toprule
 $0^\circ$ & $30^\circ$ & $60^\circ$ & $90^\circ$ \\
\midrule
\textbf{\underline{Orig.}} \\
N & N & N & N \\
8.2e-5 & 8.3e-5 & 9.2e-5 & 1.0e-4 \\
9.1e-5 & 9.3e-5 & 9.9e-5 & 1.4e-4 \\[0.6em]

\textbf{\underline{NS}} \\
N & N & N & N \\
8.1e-5 & 8.7e-5 & 9.3e-5 & 1.0e-4 \\
9.1e-6 & 9.2e-5 & 9.8e-5 & 1.4e-4 \\[0.6em]

\textbf{\underline{EW}} \\
N & N & N & N \\
8.0e-5 & 7.8e-5 & 9.1e-5 & 1.0e-4 \\
9.1e-5 & 9.2e-5 & 9.8e-5 & 1.4e-4 \\
\bottomrule
\end{tabular}
\end{subtable}

\caption{Whether net ablation is observed for four different modeled exosphere density values, as well as the maximum trough and median background deposition and ablation rates observed across the DEM. Results are included for the four modeled latitudes, and for the original (Orig.), north–south (NS) and east–west (EW) orientations. Rows labeled ``Net Abl." indicate whether net ablation is predicted by the model. ``N" indicates ablation is not observed anywhere on the map, ``T" implies ablation is observed in the trough, ``S" indicates ablation is observed in the surroundings beyond the trough, ``A" implies all facets experience ablation. For rows labeled $\Delta \mathrm{Ice_{max}}$, the maximum deposition/ablation rates recorded for the trough are given in meters. A negative value indicates net loss, or ablation of ice, while a positive number indicates net deposition. Note that if deposition dominates, the value given is the smallest value of ice deposited. Rows labeled $\Delta \mathrm{Ice_{med.}}$ give the median ice change value, in meters, across the DEM. 
}
\label{tab:ablation}
\end{table*}

\begin{figure}
    \centering
    \includegraphics[width=0.99\linewidth]{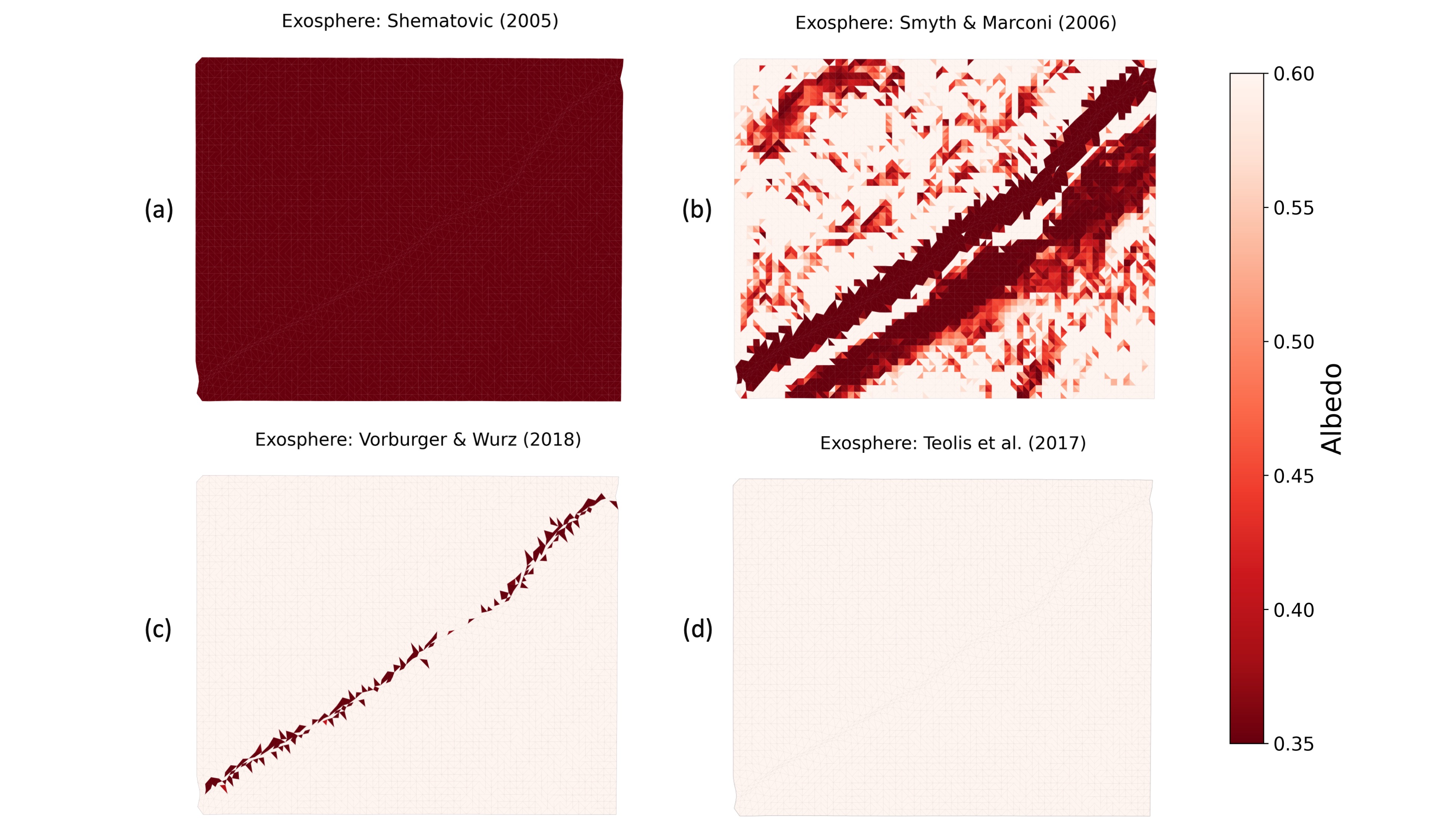}
    \caption{Albedo maps generated using model predicted lag growth rates and lag thicknesses after 100 years. Each panel shows a different calculated water exosphere density, with increasing density from panel (a) to panel (d). Light coloring means no lag layer is formed, and water icy is likely deposited. Dark red material indicates an optically thick lag layer. This layer approaches an albedo limit of about 0.35 as the optical depth of the layer grows to much greater than 1. Refer to Table \ref{tab:exovals} for specific exosphere density values. Here, (a) is the thinnest atmosphere using the lower bound value from \citet{Shematovich2005Surface-boundedEuropa}. All facets experience formation of a low albedo layer after 100 years. (b) represents the exosphere from \citet{Smyth2006EuropasImplications}, which produces diffuse low albedo growth that spreads beyond the trough. For (c),  from \citet{Vorburger2018EuropasContribution}, low albedo material is closely aligned with the trough. Lastly, (d) uses the exospheric estimate from  \citet{Teolis2017PlumePredictions} and results in net deposition of high albedo material.}
    \label{fig:albedomaps}
\end{figure}

To visualize the effect of ablation on the DEM and their potential detectability in images of Europa's surface, we use the model of section \ref{sec:albedo} to produce albedo maps including sublimation, sputtering and deposition from the exosphere. The albedo maps are shown in Figure \ref{fig:albedomaps} for the four modeled exosphere densities after after a period of 100 years. All exosphere density values considered here produce distinct low albedo patterns. The lowest modeled exosphere density quickly produces a low-albedo lag layer across the entire DEM. Exosphere values of order $10^{16}$ $\mathrm{H_2O~m^{-2}}$ result in stable low albedo material alongside bright, icy terrain. An exospheric density of 6.02 $\times 10^{16}$ $\mathrm{H_2O~m^{-2}}$ from \citet{Vorburger2018EuropasContribution} exhibits low albedo material tightly aligned with the trough. The slightly thinner exosphere of 2.3 $\times 10^{16}$ $\mathrm{H_2O~m^{-2}}$ from \citet{Smyth2006EuropasImplications} produces a low albedo lag layer in the trough and across some surroundings, especially on the warm flanks of the double ridge. In contrast, the thickest water exosphere values produce a map of deposited high albedo water ice 
(panel (d)). 

We find that ice ablation and lag formation processes proceed rapidly, with the model predicting many facets reaching a low albedo of roughly 0.35 after only tens of years. Some facets shown in Figure \ref{fig:albedomaps} would darken with additional time for models with exospheric density of order $\sim 10^{16}$ $\mathrm{H_2O~m^{-2}}$. This is especially true for the model using an exosphere density of 2.3 $\times 10^{16}$ $\mathrm{H_2O~m^{-2}}$ from \citet{Smyth2006EuropasImplications}. For this density, just over 50\% of facets develop some amount of low albedo lag in 100 years. Slightly more than 10\% of facets have reached an albedo of $\sim 0.35$ and are stable after 100 years, while about 40\% would continue to darken with more time. However, in the low and high exosphere density cases, all facets are stable as either low albedo or high albedo ice after 100 years.

\section{Discussion}

The model results show that the thermal segregation driven by low albedo lag layer formation may be a rapid process, of order $\sim 10^{0} - 10^{2}$ years in troughs if the exosphere is sufficiently thin ($< 3 \times 10^{17}~\mathrm{H_2O~m^{-2}}$) to allow for ablation. This is consistent with the fast timescales predicted by \cite{Spencer1987ThermalSatellites}. However, formation of a low volatility lag layer will reduce ablation as it thickens \citep{Spencer1987ThermalSatellites, Bramson2019AMars}. When the lag layer reaches a critical thickness, ablation will slow to near zero, effectively armoring the icy material underneath. Notably, this reduction in sublimation will not occur in a significant way until the lag layer thickness approaches or exceeds the layer reaches the critical optically thick depth. Thus, the eventual slowing of sublimation due to the lag layer will not significantly affect the darkening timescale. 

\begin{figure}
    \centering
    \includegraphics[width=0.65\linewidth]{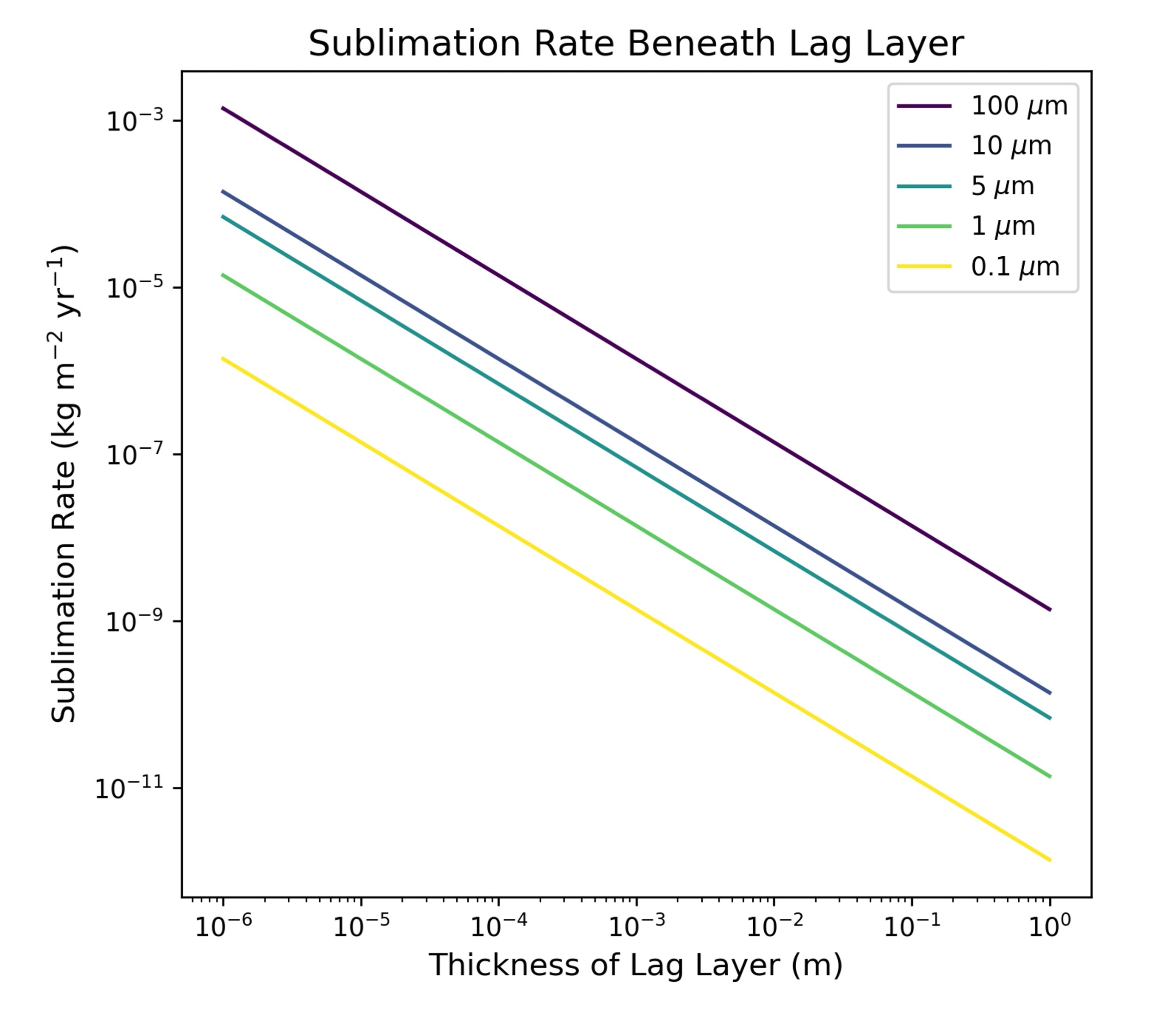}
    \caption{The annual sublimation rates of water ice beneath a growing regolith layer. Different regolith grain sizes are shown, with larger grains requiring a thicker lag layer to reduce sublimation of the ice beneath. As the layer grows the sublimation rate of the material underneath becomes negligible, especially for small regolith grains.}
    \label{fig:laggrowth}
\end{figure}

Following the relationship from \citet{Hayne2015ThermalTopography}, the sublimation rate of ice beneath a regolith layer scales roughly as $\dot E_b \sim \frac{d}{h} \dot E$. Here $d$ is the regolith grain size, h is the thickness of the regolith layer, and $\dot E$ us the sublimation rate of the material. Using the average sublimation rate of facets in the trough for $\dot E$, we estimate the lag thickness $h$ required to drop the sublimation rate to near zero. For the densest exosphere that resulted in ablation (6.02 $\times 10^{16}$ $\mathrm{H_2O~m^{-2}}$, \citet{Vorburger2018EuropasContribution}), the sublimation rate per year of ice beneath regolith is shown in Fig. \ref{fig:laggrowth}. Larger regolith particles require a thicker lag layer to reduce the sublimation rate of ice underneath. A lag depth of approximately 1 mm, forming over about $10^{3}$ years, will slow the sublimation of the material underneath to less than $10^{-5}~\mathrm{kg~m^{-2}~yr^{-1}}$ for all regolith grain sizes. Using a surface ice density of $100~\mathrm{kg~m^{-3}}$, this corresponds to a loss of about 0.01 $\mathrm{\mu m~yr^{-1}}$ of ice for the 100 $\mathrm{\mu m}$ grain size and about $10^{-10}\mathrm{~m~yr^{-1}}$ for the 1 $\mathrm{\mu m}$ grain size. If the lag layer reaches 1 m thick after just under one million years, sublimation of the ice underneath would have slowed to less than 1 mm per billion years for 1 $\mathrm{\mu m}$ grains, generally considered stable \citep{Zhang2009Cold-trappedOrigins}, and 10 mm per billion years for 100 $\mathrm{\mu m}$ grains. The speed of these timescales indicate that Europa's surface should produce a lag layer and reach a steady state quickly, with sublimation slowing to a halt after the lag layer is a few millimeters thick. As new material is deposited or geologic mechanisms produce new topography, thermal segregation will act quickly. This offers a potentially intriguing path for future study and observation, as lag material may form in some regions in less than 10 years. 

In areas like parts of the trailing hemisphere with high initial concentrations of non-ice material, ice loss may be extremely slow, and thermal segregation may not be a significant process at all. It is also worth noting that the timescales modeled in this work are orders of magnitude shorter than those predicted for dehydration of hydrated sulfates predicted by works like \cite{McCord2001ThermalConditions}.

The extent and geometry of the modeled lag layer formation varies strongly with exospheric density. As shown in Figure \ref{fig:albedomaps}, the column density used can produce an albedo map either completely covered in lag or water ice deposited by the exosphere. Some exosphere estimates result in distinct low albedo patterns. In the case of the the exosphere from \citet{Smyth2006EuropasImplications}, low albedo growth occurs in the trough, but also spreads beyond it to form significant diffuse low albedo patterns. For the slightly thicker exosphere from  \cite{Vorburger2018EuropasContribution}, low albedo material is predicted for facets within the trough. However, these are tightly aligned with the trough and no diffuse reddening is predicted outside the double ridge. At present, the model does not reproduce the splotchy diffuse reddening observed along the prominent double ridge oriented northwest to southeast in Fig. \ref{fig:doubleridges} of the Rhadamanthys terrain. Improved estimates of the water exosphere would provide critical constraints on the importance of thermal segregation on Europa's surface. 

In the case of the exospheric density from \cite{Teolis2017PlumePredictions}, net deposition dominates. Thus, it can be said that the value from \cite{Teolis2017PlumePredictions} is not consistent with our model. Unless there is a source of exospheric replenishment via a mechanism like endogenic plumes, this implies exospheric collapse. To produce consistent results, model sputtering rates would need to be increased until deposition does not dominate, and a regional balance is reached between deposition and ablation. This amount would require sputtering to increase to a value of order $10^{-5}$ meters per year to counter the deposition shown in the left panel of Figure \ref{fig:deposition}, a roughly three order of magnitude increase from the average value in \cite{Cassidy2013MagnetosphericEuropa}. The exospheric density value from \cite{Shematovich2005Surface-boundedEuropa} is also inconsistent as it predicts ubiquitous lag formation at low to middle latitudes, something which is not supported by observations. 

If a dark lag layer is formed by thermal segregation, lateral transport may distribute low albedo material beyond the initial area of concentration. This may provide a mechanism for forming some of the diffuse patterns observed on Europa's surface. Landscape evolution models that consider mass movement and transport of regolith (e.g. \cite{Richardson2009CrateringProblem} and \cite{OBrien2021PhysicalRegolith}) could serve as a tool to investigate whether low albedo lag produced by thermal segregation could reproduce observed diffuse patterns. This represents an avenue for future work. 

One important finding from this study is that a lag layer, once formed, can be a catalyst for reddening to spread in concavities where self-heating may occur. Using the previously shown albedo maps, we assign the facets that experience ablation a lower albedo value of 0.4. We feed these albedos back into the thermophysical model, resulting in higher temperatures on those facets. These facets in turn heat facets around them, causing growth of the low albedo lag layer. An example of this is shown in the supplementary materials. Redistribution of material due to impact gardening and sputtering may also spread dark material horizontally across the surface.

Topography is a critical model input, and self-heating is a significant effect in troughs. Many existing exosphere models do not consider temperatures above 132 K on Europa. With the inclusion of self-heating, our model predicts temperatures several degrees higher than those previously considered to be the subsolar peak temperatures on Europa, and at smaller scales than would have been observed. In these areas where topography results in elevated temperatures, sublimation will be a more significant process than elsewhere on the surface, provided $\mathrm{H_2O}$ ice exists on (or within $\sim1~\mathrm{mm}$) the surface. We expect this effect to be more pronounced with more detailed DEMs, as increasing roughness and topographic relief would result in greater scattering and secondary heating. Europa Clipper will obtain the high-resolution image data needed to produce more detailed topography maps over a greater fraction of Europa's surface. Consideration of sub-grid scale roughness, as well as modeling of more detailed DEMs as they become available, will be the focus of future work. 

Notably, the effect of thermal segregation could be coupled to dark material concentrations in pre-existing patterns due to Europa's geology and the global hemispheric color dichotomy. If material were deposited by plume activity (e.g. \cite{Quick2020CharacterizingEuropa}) or concentrated via frictional heating on faults (e.g. \cite{Nimmo2002Strike-slipEuropa}), the effect of thermal segregation could be enhanced. Darker material initially emplaced on the surface would lead to higher temperatures, which may extend the formation of a lag layer to much higher latitudes or to terrain with less significant self-heating than double ridges. 

In these cases, thermal segregation may further concentrate non-ice material, or it may produce or expand areas with lag layers beyond the original pattern of deposited material. Understanding the extent to which thermal processes produce dark lag layers or alter patterns from other formation mechanisms is important to constraining models for the origins of the non-ice material and the processes that are active on the Europan surface. 

Here we assume a non-ice grain size of 1 micron. Notably, hydrates on Europa may display coarser dust or grain sizes than other icy worlds like Enceladus. Previous studies have suggested grain size reaching from  ~40 µm to several hundred microns for hydrates \citep{Mishra2021AEuropa, Ligier2016VLT/SINFONICOMPOSITION}. The thickness of the lag layer scales as roughly $L \sim r_{\mathrm{p}}$, where $r_{\mathrm{p}}$ is the grain radius. Thus, as grain radius increases, the thickness required to reach optical depth will also increase in an approximately linear relationship. However, it should not have a significant effect on the timescale of lag formation until after a layer has become optically thick. It also will not change where material can ablate, as this is determined by sublimation and scattering rate, and thus by temperature. A consideration of how thermal alteration interacts with varied mixture types could offer an interesting avenue for continuing study. An intimate mixture of ice and non-ice of varied grain sizes, compositions and concentrations could produce similar darkening but highly varied photometric results.

It is important to note that sputtering is treated only approximately in our model. We do not take into account regional variations across Europa's surface \citep{Nordheim2022MagnetosphericSurface}. Given the present study's focus on local topographic effects and latitudinal variations, we use a global sputtering value, though we do include the temperature dependence of sputtering yields \citep{Cassidy2013MagnetosphericEuropa}. Extensive work has been done regarding sputtering on icy worlds, including on the angular and energetic dependence of incident ions \citep{Fama2008SputteringIons}. At present we do not distinguish between the sputtering yields on the leading and trailing hemispheres, or the incidence angle of charged particles.


\section{Conclusions}

In this work, we apply a 3D thermophysical model previously developed for airless bodies to investigate sublimation, sputtering, and dark red lag formation on Europa's double ridge troughs. Using digital elevation models, we calculate whether thermal segregation can produce the observed reddening. Assuming an initial 10\% non-ice concentration, we calculate temperatures and lag formation timescales as a result of sublimation and sputtering, and albedo maps corresponding to exosphere densities from literature. The mass balance of material on the surface is also studied, including sublimation, sputtering and deposition from the exosphere.  
 
We find that in the absence of pre-existing low albedo material, thermal segregation can produce an optically thick lag layer, similar to the reddened material observed on double ridges. There is also a strong latitudinal variation observed, with thermal segregation forming optically thick lag layers from the equatorial to mid-latitudes. However, we do not observe lag formation above 60 degrees. This suggests thermal segregation is not sufficient to produce reddening at high latitudes if low albedo material is not already present. 

Our key findings are as follows:  
\begin{itemize}
    \item We calculate temperatures in double ridges at a range of latitudes and orientations.
    \item Self-heating leads to troughs with higher peak temperatures than their surroundings and which stay warmer at night. Peak temperatures are $\sim$ 7 - 20 K higher in the trough. This results in sublimation rates in the trough that are about 10x higher at the equator and 100x higher relative to the background at $60^{\circ}$. 
    \item At middle latitudes, the maximum temperatures in the trough may be up to $\sim$ 20 K warmer than the background terrain. Topography can produce temperature contrasts that may be relevant for hot spot detection
    \item Thermal segregation can produce observable reddening from the equator to mid-latitudes in double ridge troughs.
    \item The magnitude of the effect of thermal segregation depends strongly on the density of the water exosphere, which is poorly constrained. Model predictions may be leveraged to help inform local exospheric density.
    \item If the exosphere is sufficiently thin that net ablation can occur, the warmest areas in the trough can sublimate an optically thick lag layer in $\sim$10 - 100 years.
    \item We provide predicted albedo variations for a range of exospheric water densities using lag thicknesses predicted by the model 
    \item We predict the change in the surface mass balance, including ablation or deposition, as a result of sublimation, sputtering, and a range of exosphere density values.
    \item Thermal segregation can feed back on existing low albedo regions, such as those produced by thermal segregation or by processes like plume deposition. This can result in growth of albedo contrasts between dark regions in troughs and on ridges.
\end{itemize}

The DEM modeled here includes one double ridge, and the magnitude of thermal segregation is likely to vary with the topography of the landscape and thermophysical properties. However, we find thermal segregation capable of producing reddening on a scale that may significantly affect Europa's surface. The model used in this paper is adaptable and could be applied to any available DEM of Europa's surface. 

Lag formation due to thermal segregation is likely to be enhanced with more detailed topography and roughness. At higher latitudes, increased topography or roughness may be sufficient to produce small areas of sublimation. Similarly, if pre-existing dark material is deposited, thermal segregation may be able to expand lag coverage at higher latitudes. This will be studied in future work.

Recent work has considered the effect of temperature on ice metamorphism on Europa's surface, including crystallization \citep{Mergny2025TheAlteration, Cartwright2025JWSTEuropa} and the grain size evolution caused by sintering \citep{Molaro2019TheSintering,Mergny2024ActiveTimescale}. Both processes are highly sensitive to temperature. At warmer low latitudes, crystallized ice could go through rapid re-crystallization which may outpace amorphization of ice by ions \citep{Cartwright2025JWSTEuropa}. The elevated trough surface temperatures predicted by this work may lead to the acceleration of ice sintering, which could produce larger ice grains. Though these larger grains may reduce the reflected flux scattered in the trough, they may also lower the albedo. This in turn may create a positive feedback loop that leads to higher surface temperatures and increased sublimation rates.

Most existing data of Europa’s surface is largely too resolution-limited to test these model predictions. Some of the highest resolution Galileo data may provide an opportunity for investigating the prevalence of dark material on double ridges or in other concave topography. However, this high resolution data is extremely spatially limited—imaging with $<40$ m/pixel resolution covers about 0.03\% of Europa’s surface—and concentrated on the trailing hemisphere \citep{Doggett2009GeologicSurface,Leonard2018AnalysisEuropa}. With coarser resolution it may be possible to analyze the dependence of reddening and albedo patterns on ridge orientation. Investigation of darkening as it relates to latitude is likely to require new, high resolution coverage. This analysis is left to future work.

The model results presented here will be testable upon the planned arrival of Europa Clipper to the Jupiter system in 2030. The existence and distribution of low albedo material will be critical to constrain formation of surface features. Data from E-THEMIS will elucidate the thermal environment of terrain like ridges and can be compared to model results. In addition, predictions of the thermal segregation of materials in double ridges could be used to support mapping of spectral variations with the MISE instrument.

Thermal segregation may have observable effects. Our finding of increased temperatures in troughs suggests that self-heating could be a significant mechanism driving locally elevated temperatures at mid- to high latitudes, and should be considered when interpreting data from Europa Clipper. Neglecting this effect when analyzing future E-THEMIS data may lead to false identification of thermal anomalies as endogenic hot spots. Our results indicate that, if elevated nighttime temperatures are present, they must exceed the background temperature by more than ~20 K at mid and high latitudes and ~7 K near the equator to be confidently attributed to an endogenic heat source and not to topographic effects (for a spatially resolved source). According to the thermal evolution models of hot spots by \citet{Abramov2008NumericalEuropa}, such threshold corresponds to a heat source that has remained active over the past few thousand Earth years, depending on latitude, heat source size, and surface properties. Consequently, only relatively recent endogenic activity would be detectable within Europa's ridges, as older heat sources may be masked by the effects of complex local topography. A detailed understanding of topography and the temperature contrasts that result from it will be critical when pursuing hot spot detection or identification of melted material.  

However, this work also indicates that topographical formations such as troughs may produce cold, even permanently shadowed regions at high latitudes. As sublimation should be largely ineffectual at these higher latitude regions, these areas may represent candidate sites for measuring properties like Europa's geothermal heat flow. A study of the thermal environment close to the poles could provide necessary context for upcoming observations.

\acknowledgments
The authors thank J. Spencer for valuable discussions and commentary related to this paper. Part of this work was supported by NASA's Europa Clipper project through the Precursor Science Investigations for Europa (PSIE) program under Grant No. 80NSSC25K7628. Contributions by K. C. S. were partially supported by the National Science Foundation Graduate Research Fellowship under Grant No. DGE 2040434. Any opinions, findings, and conclusions or recommendations expressed in this material are those of the author(s) and do not necessarily reflect the views of the National Science Foundation.

\section*{Conflict of Interest}
The authors declare there are no conflicts of interest for this manuscript.

\section*{Open Research Section}
The 3D thermophysical model used in this work, initially from \citet{Sorli2025AFG3}, is available at the following repository link. The version of the model used in this work, v1.0.0, is written in Python and is available on Zenodo. The DEMs of double ridges from the Rhadamanthys region used herein are also available via the same repository \citep{Sorli2025b}.

%
%

\bibliography{references}

%
%
%
%
%

\clearpage
\appendix
\section{Supplementary Material}

\begin{figure}[htp]
    \centering
    \includegraphics[width=0.75\linewidth]{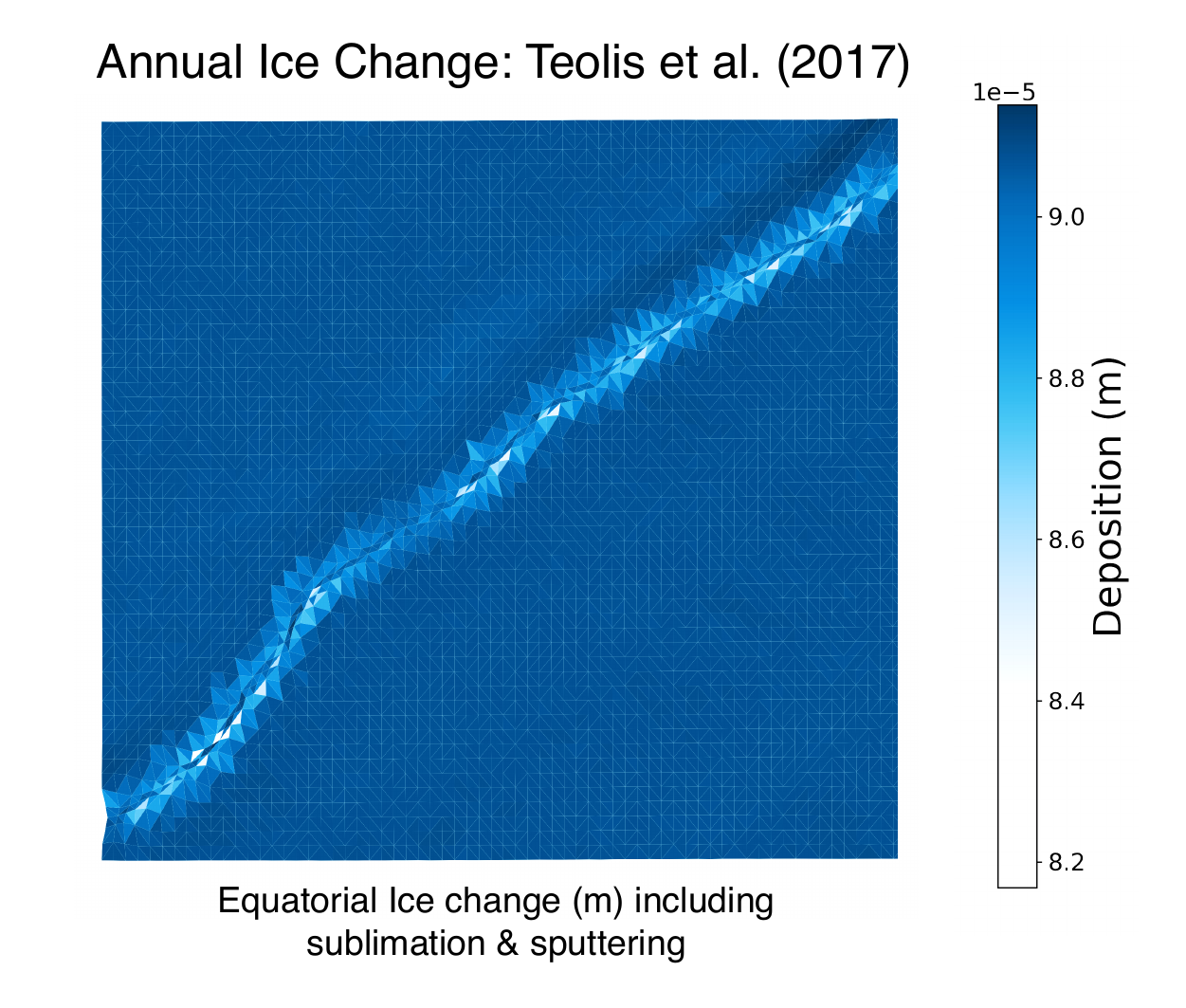} 
    \caption{The predicted annual surface mass balance change using lag thicknesses produced by the thermal model, and including the effects of sublimation, sputtering and the water exosphere. The exosphere density applied is the average of the day-night value from Teolis, Wyrick, et al (2017). Values are normalized to the range of ice change values obtained for this exospheric density to show increased detail. This exospheric density results in net deposition, shown in blue, of order $10^{-5}$ m. The trough experiences slightly less deposition, and thus appears lighter.}
    \label{fig:ap_t}
\end{figure}

\begin{figure}
    \centering
    \includegraphics[width=0.8\linewidth]{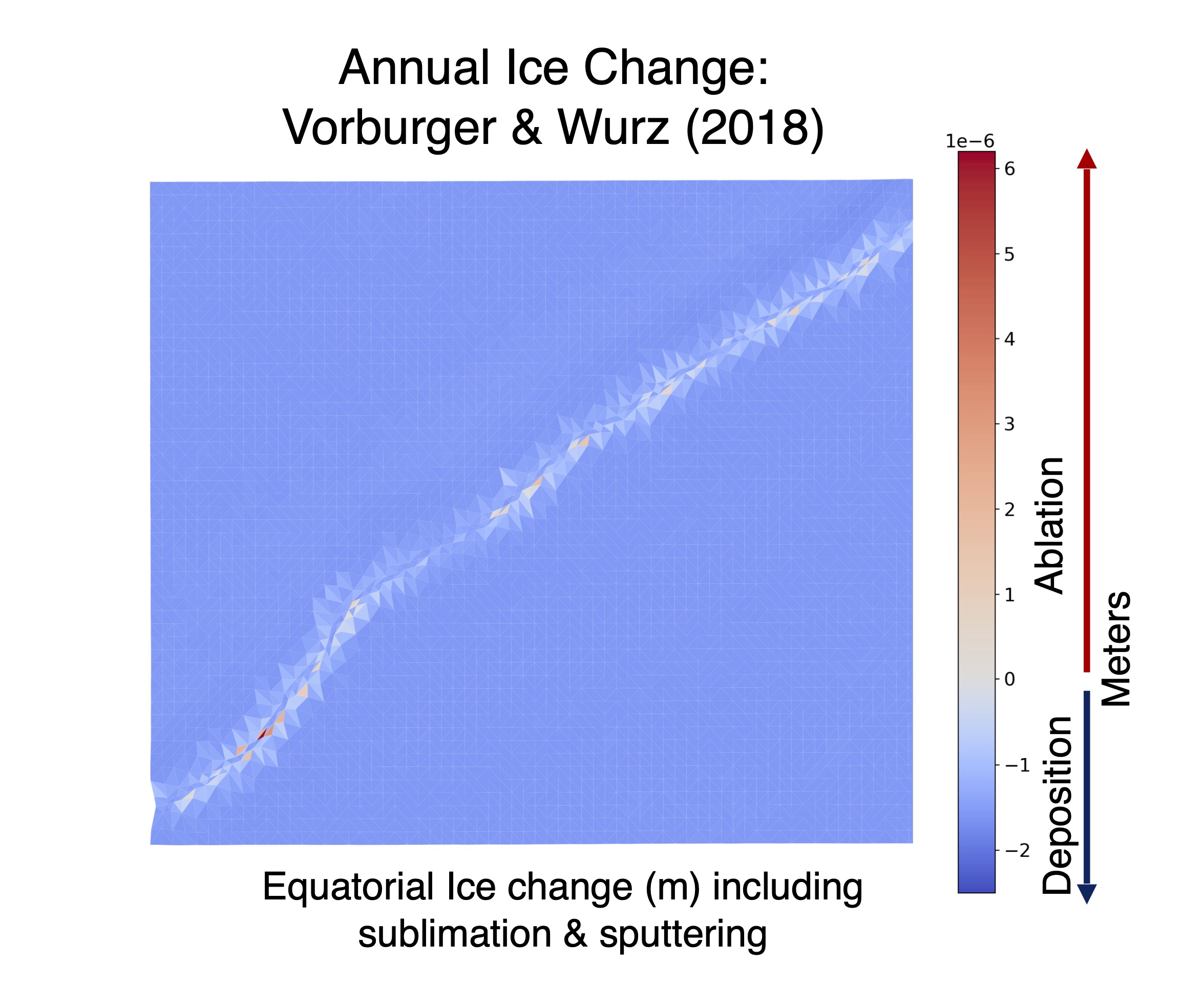}
    \caption{Expected annual surface mass balance change using lag thicknesses produced by the thermal model, including sublimation, sputtering and the water exosphere. Results pictured use the exospheric density from Vorburger \& Wurz (2018). Net ablation occurs in the trough, with white to red indicating net ablation while blue indicates net deposition. Note that the color bar is scaled in units of 1e-6 meters.}
    \label{fig:ap_VandW}
\end{figure}

\end{document}